\documentclass[aps,preprint,showpacs]{revtex4}
\usepackage{graphicx,amsmath,amssymb}
\usepackage{epsfig}
\newcommand\imagewidth{\columnwidth}

\newcommand{\rv}{{\bf r}}
\newcommand{\xv}{{\bf x}}
\newcommand{\ev}{{\bf e}}
\newcommand{\sv}{{\bf s}}
\newcommand{\Ov}{{\bf \Omega}}
\newcommand{\wv}[2]{{\bf w}_{{{v}#1}}^{#2}}
\newcommand{\nv}[2]{{\bf n}_{{v}#1}^{#2}}

\newcommand{\myext}{eps}

\newcommand{\Ovv}{{\bf \Omega'}}
\newcommand{\On}{{\bf \Omega}_N}
\newcommand{\Od}{{\bf \Omega}_D}
\newcommand{\colvec}[1]{\Bigl(\begin{smallmatrix}#1\end{smallmatrix}\Bigr)}
\newcommand{\rowvec}[1]{(#1)}

\newcommand{\Heavi}[1]{\,\Theta\!\left(#1\right)} 
\newcommand{\dirac}[1]{\delta\left(#1\right)} 

\newcommand{\abs}[1]{\left|#1\right|} 

\newcommand{\ph}{\varphi}
\newcommand{\phb}{\bar\ph}
\newcommand{\zb}{\bar z}

\newcommand{\rd}{\rds{}}   
\newcommand{\rds}[1]{\smash{R_\disk^{#1}}} 
\newcommand{\inti}{\int_0^\infty}
\newcommand{\intii}{\int_{-\infty}^\infty}

\newcommand{\kb}{k_B}
\newcommand{\kt}{\kb T}

\newcommand{\upd}{\mathrm{d}}

\newcommand{\SDD}{{SDD}}
\newcommand{\SD}{{SD}}
\newcommand{\DDD}{{DDD}}
\newcommand{\spheresphere}{{SS}}
\newcommand{\sphereneedle}{{SN}}
\newcommand{\spheredisk}{{SD}}
\newcommand{\needleneedle}{{NN}}

\newcommand{\diskneedle}{{DN}}
\newcommand{\sphere}{S}
\newcommand{\needle}{N}
\newcommand{\disk}{D}
\newcommand{\rhos}{\rho_{\sphere}}
\newcommand{\rhon}{\rho_{\needle}}
\newcommand{\rhod}{\rho_{\disk}}
\newcommand{\phifmt}[1]{\Phi_{#1}}

\begin{document}
\date{18 November 2005}

\author{Ansgar Esztermann$^1$, Hendrik Reich$^1$, and Matthias Schmidt$^{1,2}$}
\affiliation{$^1$Institut f\"ur Theoretische Physik II,
    Heinrich-Heine-Universit\"at D\"usseldorf, Universit\"atsstra\ss e 1,
    D-40225 D\"usseldorf, Germany}
\affiliation{$^2$H. H. Wills Physics Laboratory, University of Bristol,
Royal Fort, Tyndall Avenue, Bristol BS8 1TL, UK}

\title{Density functional theory for colloidal mixtures of \\
       hard platelets, rods, and spheres}

\begin{abstract}
A geometry-based density functional theory is presented for mixtures
of hard spheres, hard needles and hard platelets; both the needles and
the platelets are taken to be of vanishing thickness. Geometrical
weight functions that are characteristic for each species are given
and it is shown how convolutions of pairs of weight functions recover
each Mayer bond of the ternary mixture and hence ensure the correct
second virial expansion of the excess free energy functional. The case
of sphere-platelet overlap relies on the same approximation as does
Rosenfeld's functional for strictly two-dimensional hard disks. We
explicitly control contributions to the excess free energy that are of
third order in density. Analytic expressions relevant for the
application of the theory to states with planar translational and
cylindrical rotational symmetry, e.g.\ to describe behavior at planar
smooth walls, are given. For binary sphere-platelet mixtures, in the
appropriate limit of small platelet densities, the theory differs from
that used in a recent treatment [L. Harnau and S. Dietrich,
Phys. Rev. E {\bf 71}, 011504 (2004)]. As a test case of our approach
we consider the isotropic-nematic bulk transition of pure hard
platelets, which we find to be weakly first order, with values for the
coexistence densities and the nematic order parameter that compare
well with simulation results.
\end{abstract}


\maketitle

\section{Introduction}
Dispersions of non-spherical colloidal particles are model systems to
study a broad range of phenomena in condensed matter ranging from
fluid phase separation to the emergence of liquid crystalline
ordering. Examples of such systems are clay suspensions
\cite{dijkstra95clay,dijkstra97clay}, dispersed gibbsite platelets
\cite{vanderbeek05}, mixtures of silica spheres and silica-coated
boehmite rods \cite{oversteegen04sphereRod}, wax disks \cite{mason02},
or nonaqueous suspensions of laponite and montmorillonite
\cite{leach05}. While already pure systems often possess complex
liquid crystal phase behavior, binary mixtures additionally may demix
into bulk phases with different chemical compositions, e.g.\ like in
dispersions of disks and spheres \cite{mason02}.
Such mixtures are often viewed as composed of a primary species (here
the spheres) that interact with an effective depletion potential that
is generated by the second component referred to as the depletion
agent (here the disks). The depletion interaction is primarily
attractive with a range of attraction similar to the size of the
depletant agent and with a strength that is ruled by the concentration
of the depletion agent. Under appropriate conditions the depletion
interaction may be sufficiently strong in order to drive a phase
transition that is anaologous to the gas-liquid phase separation in
simple substances, and that manifests fluid-fluid demixing when
regarding the full mixture.

Recent theoretical work has been devoted to fluids of platelike
particles near a hard wall \cite{harnau02wall}, an interaction site
model for lamellar colloids \cite{harnau01interactionSite}, wetting
and capillary nematization of binary hard-platelet and hard-rod fluids
\cite{harnau02nematization}, the lamellar Zwanzig model
\cite{harnau02zwanzig,bier04binaryPlatelet}, and colloidal hard-rod
fluids near geometrically structured substrates \cite{harnau04}.  The
phase diagram of mixtures of hard colloidal spheres and discs was
obtained within a free-volume scaled particle approach
\cite{oversteegen04spt}. The depletion potential between two spheres
immersed in a sea of platelets was studied in detail
\cite{oversteegen04accuracy,harnau04depletion}, and found to compare
well with predictions from the Derjaguin approximation
\cite{oversteegen04accuracy}.  Sedimentation was found to influence
liquid crystal phase transitions of colloidal platelets
\cite{vanderbeek04}, as well as multi-phase equilibria in mixtures of
platelets and ideal polymer \cite{wensink04}.

The fundamental-measure theory (FMT) is an (approximate)
density-functional theory (DFT) \cite{evans79,evans92}, originally
proposed by Rosenfeld for additive hard sphere mixtures
\cite{rosenfeld89}.  An early extension to convex non-spherical
particles has been given in
Refs.~\cite{rosenfeld94convex,rosenfeld95convex}.  When applied to
homogeneous and isotropic fluid states, this theory yields the correct
second virial coefficient of the equation of state, but fails to
recover the exact density functional up to second order in density. To
remedy the latter problem, an interpolation between the hard sphere
Rosenfeld functional and the Onsager functional for elongated rods was
proposed \cite{cinacchi02} and applied to the bulk isotropic-nematic
transition. An FMT that originated from the treatment of parellel hard
cubes \cite{Cuesta96,Cuesta97prl,Cuesta97jcp} was also used very
successfully for the Zwanzig model, see e.g.\
\cite{martinezraton02,martinezraton03}. See also Refs.\
\cite{chamoux96,chamoux98,perera04demixing} for the treatment of
various hard body system mixtures. Recently attention was paid to
binary mixtures of hard rods and polymers
\cite{bryk04,bryk05rodsPolymer}. Also notable is the treatment of hard
body fluids based on two-point measures
\cite{wertheim94I,wertheim96II,wertheim96III}.

The Bolhuis-Frenkel model of hard spheres and vanishingly thin hard
needles \cite{bolhuis94} can be considered as the simplest model hard
core mixture of spheres and rods. Previous work was concerned with the
formulation of a DFT for this model \cite{schmidt01rsf}, and an
extension to include rod-rod interaction on the Onsager (second
virial) level \cite{brader02rsa,esztermann04rsc}. Predictions for the
orientation ordering of the rods at a free interface between
(isotropic) sphere-rich and sphere-poor phases \cite{brader02rsa} were
successfully confirmed by simulations \cite{bolhuis03rsb}. Adding a
third component to this system, the fluid demixing phase behavior of
ternary mixtures of spheres, rods, and model polymers was found to be
rich \cite{schmidt02cpn}. Quenching one of the components led to
investigations of hard spheres immersed in random rod networks
\cite{schmidt03porsn}, and of the isotropic-nematic transition of rods
in matrices of immobilized spheres \cite{schmidt04poro}. The present
paper is concerned with the introduction of hard platelets to the FMT
framework.

Central to FMT is the so-called deconvolution of the Mayer bond
$f_{ij}$ between particles of species $i$ and $j$ into ``weight
functions'' $w_i^\nu$ that are characteristic of the shape of
particles of each species $i$; the label $\nu$ indicates the type of
weight function. Recall that the Mayer bond is related to the pair
interaction potential $V_{ij}$ between species $i$ and $j$ via
$f_{ij}=\exp(-\beta V_{ij})-1$, where $\beta=1/(\kt)$, $\kb$ is the
Boltzmann constant and $T$ is temperature.
%
%
For hard bodies $f_{ij}=-1$ if the two particles overlap and
$f_{ij}=0$ otherwise. In the FMT framework a convolution integral is
used to express $f_{ij}$ as a sum of terms each of which is bilinear
in the weight functions, i.e.\ possesses the structure $w_i^\nu\ast
w_j^\tau$, where $i, j$ label the species, $\nu, \tau$ label the type
of weight function, and the asterisk indicates the (three-dimensional)
convolution. The weight functions are characteristic of the geometry
of single particles, i.e.\ vanish (already) beyond the physical extent
of {\em one} particle. Note that this is different from the behavior
of $f_{ij}$, which extends to larger separations and vanishes provided
{\em two} particles do not overlap. However the convolution of pairs
of (shorter ranged) weight functions restores the correct (longer
ranged) non-locality of the Mayer bond. This concept was originally
introduced for hard sphere mixtures \cite{rosenfeld89}.  An extension
to arbitrarily shaped convex bodies was proposed in Refs.\
\cite{rosenfeld94convex,rosenfeld95convex}, which yields the correct
second virial coefficients, but gives only an approximation for the
Mayer bonds. This is insufficient to describe e.g.\ nematic ordering
(see e.g.\ Ref.\ \cite{cinacchi02} for a discussion). However,
obtaining exact deconvolutions is possible in principle, as the cases
of the Mayer bond between a sphere and a vanishingly thin needle
\cite{schmidt01rsf} and between a pair of thin rods \cite{brader02rsa}
demonstrate.

To give an overview of the general structure of FMT, the weight
functions $w_i^\tau$ are used to to build so-called weighted densities
via convolution with the bare one-body density distribution (of each
species in the case of a mixture). The weighted densities are then
input as arguments to an (analytically given) excess free energy
density, $\Phi$. Integrating $\Phi$ over space (and director space in
case of anisotropic particles) yields the excess free energy of the
system, which is the only unknown contribution to the grand potential
functional being the central quantity in (classical) DFT. For any
external potential (like describing gravity or a wall) minimizing the
grand potential with respect to the one-body density distribution(s)
gives both the value of the grand potential and the structural
information contained in the density profile(s).


Here we show how to treat platelet-shaped particles and mixtures with
both spheres and needles within the same framework. The relevant
weight functions are given and it is shown explicitly how all Mayer
bonds for the ternary mixture are obtained through convolutions. For
two-dimensional hard disks the Rosenfeld functional only yields an
approximation for the exact Mayer bond. Despite this deficieny the
two-dimensional Rosenfeld functional is considered to be a useful tool
to study inhomogeneous situations, see e.g.\ Ref.\ \cite{rasmussen02}
for an investigation of laser-induced freezing and melting of confined
colloidal particles.  The treatment of the sphere-platelet case
inevitably leads to the same defect. We do achieve, however, exact
deconvolutions of the Mayer bonds for platelet-platelet and
platelet-needle interactions.

Furthermore we treat explicitly contributions to the excess free
energy functional that are of third order in densities. While the
treatment of second order terms was a crucial input for the original
construction of FMT for hard spheres \cite{rosenfeld89}, this route
has been only recently pursued in detail \cite{cuesta02}, and the
authors conclude that they come ``close to the edge of FMT'' in
attempting to incorporate properties of the exact expansion of the
free energy functional in density on the third virial level into the
approximate framework of FMT. In particular these delicate cases
appear when particle touch, i.e.\ configurations that are close to the
edge of the particles.

Hence we approximate the exact third virial level (the triangle in the
diagrammatic Mayer $f_{ij}$-expansion) by terms that are non-vanishing
only for cases with common triple intersection of the three particles
involved.  Global prefactors are used to compensate for the ``lost
cases'' \cite{tarazona97,cuesta02} in order to yield reasonable third
virial coefficients.

As a test case for the accuracy of the theory we consider the pure
system of hard platelets. This is known to undergo a weak first order
isotropic-nematic phase transition \cite{frenkel82}. In contrast to
the application to thin rods, the Onsager (second virial) functional
for platelets yields only a qualitatively correct account of the phase
transition: The transition density, density jump, and value of the
nematic order parameter in the coexisting nematic are significantly
overestimated. Our theory has the same (exact) second order
contribution to the free energy, but also sports a further third-order
term. Higher than third-order terms are absent due to the
scaled-particle roots \cite{barker76} of the approach; the vanishing
thickness and hence vanishing volume of the platelets effectively
truncates the series. Our results show that FMT considerably improves
on the Onsager treatment, and yields a quantitatively correct picture,
albeit with slightly too small transitition densities.

Harnau and Dietrich recently considered bulk and wetting phenomena in
a binary mixture of colloidal hard spheres and hard platelets
\cite{harnau04spherePlate}, an interesting limiting case of the
present ternary mixture. We postpone a detailed discussion of the
relationship of the current work to Ref.\ \cite{harnau04spherePlate}
to Sec.\ \ref{SECplateconclusions}.

The paper is organized as follows. In Sec.\ \ref{SECplatemodel} we
define the model of a ternary hard body mixture of spheres, needles,
and platelets. In Sec.\ \ref{SECdeconvolution} the Mayer bonds are
represented as convolutions of (weight) functions characteristic for
single particles. In Sec.\ \ref{SECtriangle} we proceed to control the
third virial level. Readers who are interested in the structure of the
resulting theory rather than its derivation might wish to skip Secs.\
\ref{SECdeconvolution} and \ref{SECtriangle} and directly go to Sec.\
\ref{SECweightedDensities} where the weighted densities are defined as
convolutions of the weight functions with the bare density
profiles. Sec.\ \ref{SECdft} presents the excess free energy
functional.  Explicit expressions for the relevant quantities in
planar and uniaxial geometry are given in Sec.\
\ref{SECplanarGeometry}.  We apply the theory to the isotropic-nematic
transition of hard platelets in Sec.\ \ref{SECresults}, and conclude
in Sec.\ \ref{SECplateconclusions}.

\section{Model}
\label{SECplatemodel}
We consider a mixture of hard spheres (species $S$) with diameter
$\sigma=2R$, where $R$ is the radius, hard platelets (species $D$) of
diameter $2\rd$ and vanishing thickness, and hard needle-like rods
(species $N$) of length $L$ and vanishing thickness. The pair
interaction potential $V_{ij}$ between any two particles of species
$i,j=S,D,N$ is infinite if their geometrical shapes overlap and zero
otherwise. The one-body density distribution of species $i=S,D,N$ is
denoted by $\rhos(\rv)$, $\rhod(\rv,\Ov)$, and $\rhon(\rv,\Ov)$,
respectively, where $\rv$ is the position of the particle center, and
$\Ov$ is a unit vector pointing along (normal to) the shape of the
needle (platelet) describing the particle orientation in space.

\section{Deconvolution of the Mayer bonds}
\label{SECdeconvolution}
\subsection{The sphere-sphere and sphere-needle Mayer bonds}
\label{SECmayerSSandSN}
For completeness we first summarize results from the literature for
hard spheres and their mixtures with needles. Rosenfeld's hard sphere
weight functions \cite{rosenfeld89} are
\begin{align}
  w_3^{\sphere}(\rv) &= \Theta(R-|\rv|), \label{EQw3S}\\
  w_2^\sphere(\rv) &= \delta(R-|\rv|),\label{EQw2S}\\
  \wv{2}{\sphere}(\rv) &= w_2^\sphere(\rv) \frac{\rv}{|\rv|}, \label{EQwv2S}
\end{align}
where $\Theta(\cdot)$ is the unit step (Heaviside) function and
$\delta(\cdot)$ is the Dirac distribution. Here and in the following
the $w_\tau^i$ are quantities with the dimension of $({\rm
length})^{\tau-3}$; the subscript $v$ indicates a vectorial quantity.
Further, linearly dependent, weight functions are
$w_1^\sphere(\rv)=w_2^\sphere(\rv)/(4\pi R)$,
$w_0^\sphere(\rv)=w_2^\sphere(\rv)/(4\pi R^2)$, and
$\wv{1}{\sphere}(\rv)=\wv{2}{\sphere}(\rv)/(4\pi R)$. For pure hard
spheres the Mayer bond is obtained through
\begin{align}
  -f_{\spheresphere}(\rv)/2 = & w_3^\sphere(\rv) \ast w_0^\sphere(\rv) + 
    w_2^\sphere(\rv) \ast w_1^\sphere(\rv) \nonumber\\
    &\quad - \wv{2}{\sphere}(\rv) \ast \wv{1}{\sphere}(\rv),
    \label{EQfSS}
\end{align}
where the (three-dimensional) convolution is defined as 
\begin{equation*}
  h(\rv)\ast g(\rv)=\int \upd^3x\, h(\xv) g(\xv-\rv),
\end{equation*}
and also implies a scalar product between vectors, as appears in the
last term on the right hand side of Eq.\ (\ref{EQfSS}).

A pair of vanishingly thin needles does not experience excluded volume
interactions, but needles do interact with hard spheres. For such a
binary mixture the needle weight functions used in Ref.\
\cite{schmidt01rsf} were obtained from the prescription of Refs.\
\cite{rosenfeld94convex,rosenfeld95convex}, and are given by
\begin{align}
  w_0^\needle(\rv,\Ov) &= 
   \frac{1}{2}\left[\delta\left(\rv-\frac{L}{2}\Ov\right)
       +\delta\left(\rv+\frac{L}{2}\Ov\right)\right],
   \label{EQw0N}\\
  w_1^\needle(\rv,\Ov) &= \frac{1}{4}\int_{-L/2}^{L/2} \upd l\,\delta(\rv - l\Ov).
   \label{EQw1N}
\end{align}
Introducing a ``mixed'' weight function that is non-vanishing on 
the surface of a sphere, but carries a dependence on orientation,
\begin{align}
  w_2^{\sphereneedle}(\rv,\Ov) &= 2 |\wv{2}{\sphere}(\rv)\cdot\Ov|,
  \label{EQw2SN}
\end{align}
allows one to obtain the Mayer bond between sphere and needle through
convolutions,
\begin{align}
 -f_{\sphereneedle}(\rv,\Ov) = w_3^\sphere(\rv) \ast w_0^\needle(\rv,\Ov)
 + w_2^{\sphereneedle}(\rv,\Ov) \ast w_1^\needle(\rv,\Ov),
 \label{EQfSN}
\end{align}
see Ref.\ \cite{brader02rsa} for the explicit calculation. For the
case of residual rod-rod interactions in the Onsager limit,
$f_{\needleneedle}$ can also be deconvolved into weight functions, see
Ref.\ \cite{brader02rsa,esztermann04rsc} for the details. In the
following, however, we will restrict ourselves to the case of
vanishingly thin needles.

\subsection{Strictly two-dimensional hard disks}
\label{SECtwod}
As a prerequisite for our subsequent treatment of platelets in three
dimensions, we give an overview of Rosenfeld's functional for the
model of strictly two-dimensional disks. This constitues a
prerequisite for considering the overlap between a hard sphere and a
hard platelet in three dimensions (Sec.\ \ref{SECmayerSD} below). Here
we deal with a multicomponent two-dimensional mixture of disks with
radii $R_i$ of species $i$, and characterized by the pair potential
$V_{ij}(r)=\infty$ if $r<R_i+R_j$ and zero otherwise, where $r$ is the
center-center distance between the disks of species $i$ and $j$. (When
regarded from the viewpoint of the three-dimensional model, all disks
possess the same orienation $\Ov$, perpendicular to the
two-dimensional plane of position coordinates; we will not delve into
this delicate dimensional crossover in the following). Rosenfeld's
weight functions for this model are
\begin{align}
  w_2^{(i)}(\rv) &= \Theta(R_i-|\rv|),\\
  w_1^{(i)}(\rv) &= \delta(R_i-|\rv|),\\
  \wv{1}{(i)}(\rv) &= w_2^{(i)}(\rv) \frac{\rv}{|\rv|},
  \label{EQwv1twod}
\end{align}
and there is an additional linearly dependent weight function,
$w_0^{(i)}(\rv) = w_2^{(i)}(\rv)/(2\pi R_i)$.  The exact Mayer bond
$f_{ij}$ between species $i$ and $j$ is then approximated through
$f_{ij}\approx f_{ij}^*$, with
\begin{equation}
  \begin{split}
  -f_{ij}^*(r) &=  w_0^{(i)}(\rv) \ast w_2^{(j)}(\rv) +
         w_2^{(i)}(\rv) \ast w_0^{(j)}(\rv) \\  
       &\qquad + 
         \frac{1}{2\pi}\Bigl( w_1^{(i)}(\rv) \ast w_1^{(j)}(\rv)
          -\wv{1}{(i)}(\rv) \ast \wv{1}{(j)}(\rv) \Bigr),
  \end{split}
  \label{EQmayerTwoD}
\end{equation}
where here (and only here) $\ast$ denotes the two-dimensional
convolution, 
\begin{equation*}
  h(\rv)\ast g(\rv)=\int\upd^2x\, h(\xv) g(\xv-\rv),
\end{equation*}
and again the convolution implies a scalar product between vectors.

\begin{figure}
  \includegraphics[width=\imagewidth]{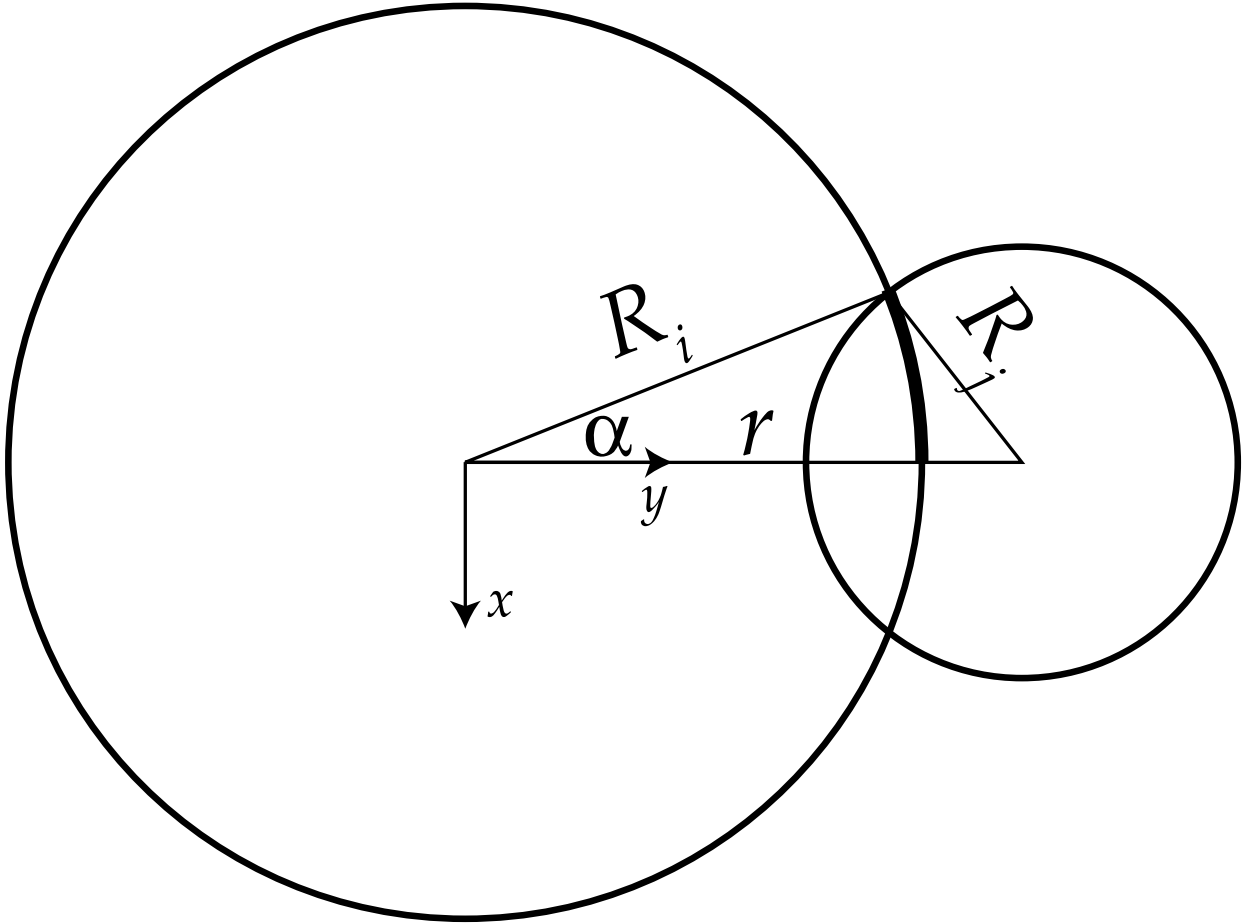}
  \caption{Overlapping pair of strictly two-dimensional hard disks
    with radii $R_i$ and $R_j$ and center-center distance $r$. Shown
    are the $x$ and $y$ axis of the coordinate system used, the angle
    $\alpha$ between the y-axis and $R_i$, as well as half of the part
    (bold) of the circumference of disk $i$ that lies inside disk
    $j$.}
  \label{fig:2darc}
\end{figure}

As our subsequent treatment of the three-dimensional sphere-platelet
overlap relies heavily on Eq.\ (\ref{EQmayerTwoD}), we calculate the
explicit functional dependence of $f_{ij}^\ast(r)$ on $r$ and discuss
some of its properties. We choose two-dimensional Cartesian
coordinates $\rv$ and $\xv$, such that $\rv=\rowvec{0,r}$ and
$\xv=\rowvec{x'\sin\ph',x'\cos\ph'}$, which allows us to write the
first term on the r.h.s.\ of Eq.\ (\ref{EQmayerTwoD}) as
\begin{align}
  \int\upd^2x\, w_0^{(i)}(\xv)&w_2^{(j)}(\xv-\rv)\nonumber\\
    &=\int_0^{2\pi}\upd \ph'\inti x'\,\upd x'\frac{1}{2\pi R_i}\delta(x'-R_i)
      \Heavi{R_j-\sqrt{x'^2+r^2-2rx'\cos\ph'}}\\
    &=\frac{1}{\pi}\arccos\Bigl(\frac{r^2+R_i^2-R_j^2}{2rR_i}\Bigr)
      \:\Heavi{2rR_i-\abs{r^2+R_i^2-R_j^2}},
      \label{eq:2d:arc}
\end{align}
which is, by the cosine theorem, equal (up to a factor of $2\pi$) to
the length of the arc which the rim of the disc $i$ traces across the
interior of disc $j$ (see Fig.~\ref{fig:2darc}). By symmetry, the
second term on the r.h.s.\ of Eq.\ (\ref{EQmayerTwoD}) gives Eq.\
(\ref{eq:2d:arc}) with $R_i$ and $R_j$ interchanged.  To calculate the
remaining third term on the r.h.s.\ of Eq.\ (\ref{EQmayerTwoD}), we
use the same setup as above and obtain
\begin{align}
  \frac{1}{2\pi}\int\upd^2x\, \Bigl(
  w_1^{(i)}(\xv)&w_1^{(j)}(\xv-\rv)-
  \wv{1}{(i)}(\xv)\cdot\wv{1}{(j)}(\xv-\rv) \Bigr) \nonumber\\
  \begin{split}
    &=\frac{1}{2\pi}\int_0^{2\pi}\upd\ph'\!\!\!\inti \upd x'\,x'\dirac{R_i-x'}
      \dirac{R_j-\sqrt{r^2+x'^2-2rx'\cos\ph'}}\\
    &\qquad\times\left(1-\frac{1}{R_i R_j}
        \colvec{x'\sin\ph'\\x'\cos\ph'}\cdot
        \colvec{x'\sin\ph'\\x'\cos\ph'-r}\right)
   \end{split}\\
    &=\frac{1}{\pi} \Heavi{R_i-|R_j-r|}
        \frac{r^2-(R_i-R_j)^2}
          {\sqrt{4r^2R_i^2-(r^2+R_i^2-R_j^2)^2}}. \label{eq:2d:cusp}
\end{align}
In the special case that both platelets have the same radius,
$R_i=R_j$, as relevant for describing the pure hard disk system, Eq.\
(\ref{eq:2d:cusp}) further simplifies to
\begin{equation}
  \frac{1}{2\pi}\int\upd^2x\,
  \Bigl( w_1^{(i)}(\xv)w_1^{(i)}(\xv-\rv)-
  \wv{1}{(i)}(\xv)\cdot \wv{1}{(i)}(\xv-\rv)\Bigr)
    =\Heavi{2R_i-r}
        \frac{r}{\pi\sqrt{4R_i^2-r^2}}.
\end{equation}

\begin{figure}
  \includegraphics[width=\imagewidth]{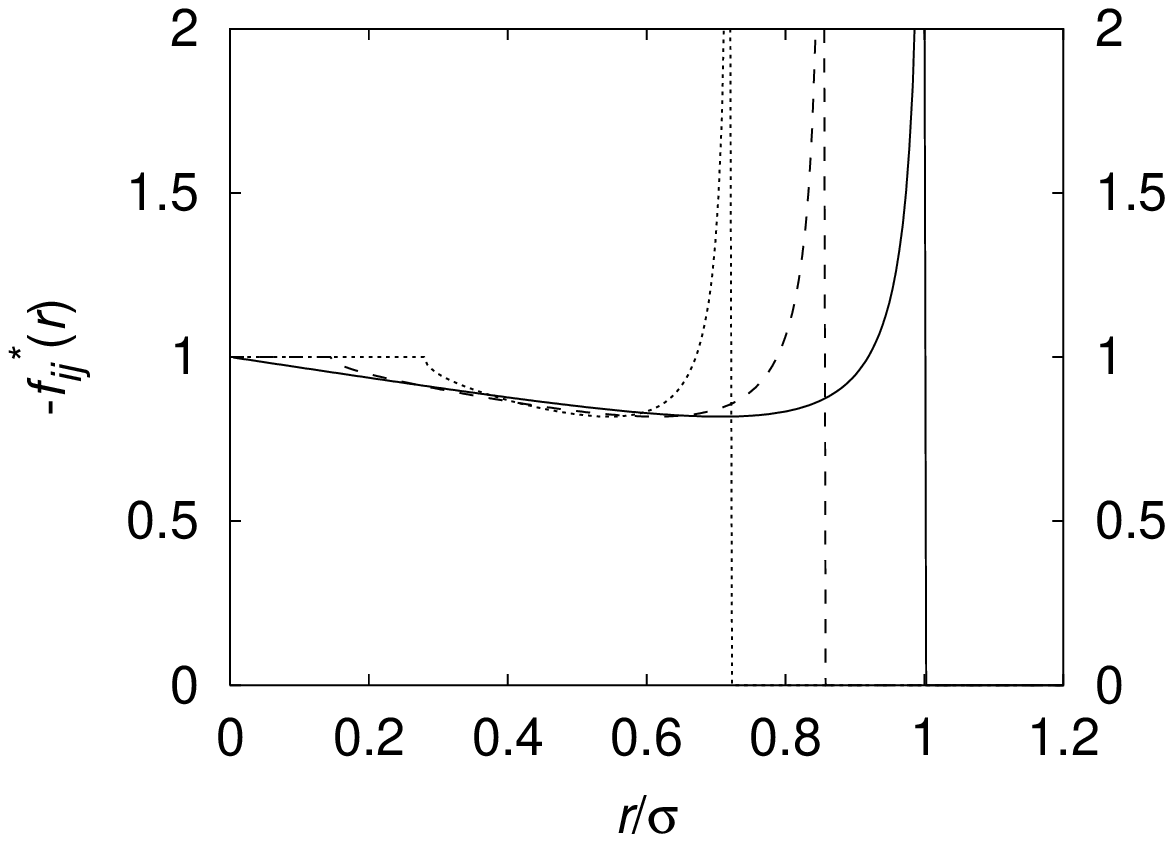}
  \caption{Rosenfeld's approximation for the (negative) Mayer bond,
    $-f_{ij}^\ast(r)$, of two-dimensional hard disks of radii $R_i$
    and $R_j$, for size ratios $R_i/R_j= 0.44, 0.71, 1$ (from left to
    right) as a function of the (scaled) center-to-center distance
    $r/\sigma$, where $\sigma=2R_j$. In Sec.\ \ref{SECmayerSD} the
    same result is obtained for the Mayer bond in three dimensions
    between a sphere of radius $R$ and a platelet of radius $R_D$ as a
    function of the (scaled) distance $r/\sigma$ between the center of
    the platelet and the axis of the sphere (parallel to the
    orientation $\Ov$ of the platelet), where $\sigma=2\rd$. The
    (scaled) distance between the center of the sphere and the plane
    of the platelet is $0, 0.7, 0.9$ (right to left), and $R=R_D$. The
    result for the Mayer bond obtained is exact in cases (not shown)
    where the platelet i) fully cuts through the sphere or ii) lies
    completely inside the sphere (provided $R_D<R$).}
  \label{FIG2d}
\end{figure}

We plot $f_{ij}^\ast(r)$ in Fig.\ \ref{FIG2d} for three typical values
of the size ratio $R_i/R_j=0.44,0.71,1$. For small separations $r$,
such that the (small) disk $i$ lies completely inside the (big) disk
$j$, the result is exact, $f_{ij}^\ast(r)=-1$ for $r<R_j-R_i$. As $r$
increases, $-f_{ij}^\ast$ first decreases and then increases,
exhibiting a divergence at contact, $r=R_i+R_j$. Beyond this
threshold, for $r>R_i+R_j$, $f_{ij}^\ast$ is (correctly) indentically
zero. Note that the deviations around the exact value inside the core,
$f_{ij}=-1$, balance such that the second virial coefficient is exact,
$\int d^2r f_{ij}(r) = \int d^2r f_{ij}^\ast(r)=-\pi(R_i+R_j)^2$.

\subsection{Platelet-platelet Mayer bond in three dimensions}
\label{SECplateletPlateletMayerBond}
Although the particle shapes appear to be equivalent, overlapping pair
configurations are very different for freely rotating platelets in
three dimensions as compared to the strictly two-dimensional case.
Note that in three dimensions we only need to consider cases of
differing orientations of both platelets, $\Ov\neq\Ov'$, as
configurations with strictly equal orientations, $\Ov=\Ov'$, carry
vanishing statistical weight. The intersection volume differs markedly
from the two-dimensional case (area of intersection), requiring very
different weight functions to express the Mayer bond. This task can be
accomplished exactly, in contrast to the two-dimensional case
above. The direct application of Rosenfeld's recipe
\cite{rosenfeld94convex,rosenfeld95convex} to platelets yields the
scalar weight functions
\begin{align}
  w_2^D(\rv,\Ov) &= 2\Theta(\rd-|\rv|)\delta(\rv\cdot\Ov),
  \label{EQw2DRosenfeld}\\
  w_1^D(\rv,\Ov) &= \delta(\rd-|\rv|)\delta(\rv\cdot\Ov)/8,
  \label{EQw1DRosenfeld}\\
  w_0^D(\rv,\Ov) &= \frac{1}{2\pi \rd}\delta(\rd-|\rv|)\delta(\rv\cdot\Ov),
  \label{EQw0DRosenfeld}
\end{align}
corresponding to the surface, $\xi_2^D = 2\pi \rds{2}$, integral
mean curvature, $\xi_1^D= \pi \rd/4$, and Euler characteristic,
$\xi_0^D=1$, of the platelets, respectively.  The weight function
that describes the particle volume vanishes, $w_3^\sphere(\rv,\Ov)=0$,
due to the vanishing thickness, and hence vanishing volume of the
platelets, $\xi_3^D=0$.

\begin{figure}
  \includegraphics[width=\imagewidth]{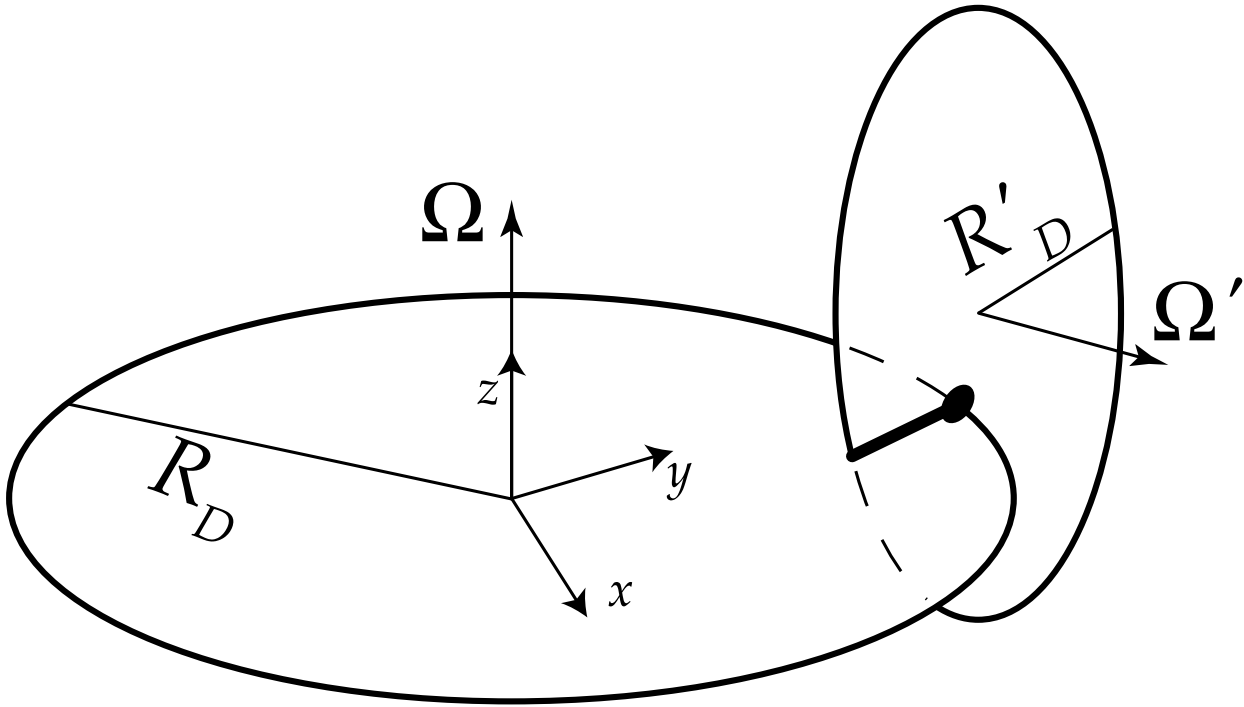}
  \caption{ Setup for the calculation of the platelet--platelet Mayer
    bond.  Platelet $D$ is located in the $x$--$y$ plane with its
    center at the origin.  Platelet $D'$ has its center in the
    $y$--$z$ plane and an arbitrary orientation.  The calculation
    checks for the intersection of the rim of $D$ and the interior of
    $D'$ (bold dot).  The one-dimensional intersection volume of the
    two platelets is shown as a bold line.  }
  \label{fig:dd}
\end{figure}

In contrast to Rosenfeld's treatment
\cite{rosenfeld94convex,rosenfeld95convex}, we here aim at an exact
deconvolution of $f_{DD}(\rv,\Ov',\Ov)$, where $\rv$ is the
center-center distance vector between both particles and $\Ov$ and
$\Ov'$ are their orientations. We introduce
\begin{align}
  w_1^{DD}(\rv,\Ov;\Ov') = & 
  \frac{2}{\rd}|\Ov \cdot (\Ov'\times \rv)| w_1^D(\rv,\Ov),
 \label{EQw1DD}
\end{align}
where $w_1^D(\rv,\Ov)$, given through Eq.\ (\ref{EQw1DRosenfeld}),
describes the platelet rim, and $\Ov'$ is the orientation of the
second platelet. Keeping Rosenfeld's surface weight function,
$w_2^D(\rv,\Ov)$ as given in Eq.\ (\ref{EQw2DRosenfeld}), we
recover the Mayer bond between platelets via convolution,
\begin{align}
  -f_{DD}(\rv,\Ov;\Ov') = & \;
    w_1^{DD}(\rv,\Ov;\Ov') \ast w_2^D(\rv,\Ov')\nonumber\\ &
   \quad +w_2^D(\rv,\Ov) \ast w_1^{DD}(\rv,\Ov';\Ov).
   \label{EQfDD}
\end{align}
As the r.h.s.\ of Eq.\ (\ref{EQfDD}) consists of two symmetric terms,
it is sufficient to consider the first one; the second one gives an
equivalent contribution to $f_{DD}$. Without loss of generality
we place the particles such that one platelet is located at the origin
with its orientation vector pointing up (into the positive
$z$-direction), and place the 
center of the
other platelet in the $y$--$z$ plane;
see Fig.~\ref{fig:dd} for an illustration. The chosen coordinates are:
$\rv=(0,r,z)$, $\Ov=(0,0,1)$, $\Ovv=(\sqrt{1-\zb^2}\sin\phb$,
$\sqrt{1-\zb^2}\cos\phb,\zb)$, and the integration variable is
$\xv=(r'\sin\ph',r'\cos\ph',z')$.
Using (\ref{EQw1DRosenfeld}) and (\ref{EQw1DD}) we insert the explicit
form $w_1^{DD}(\rv,\Ov;\Ovv) = \abs{\Ovv\cdot(\rv\times\Ov)}
\dirac{\rd-\abs{\rv}} \dirac{\rv\cdot\Ov}/(8\rd)$ into the first term
on the r.h.s.\ of (\ref{EQfDD}) to obtain
\begin{align}
  \int \upd^3x\,&w_1^{DD}(\xv,\Ov;\Ovv)
      w_2^D(\xv-\rv,\Ovv)\nonumber\\
    \begin{split}
    &=\frac{1}{2\rd}\int_0^{2\pi}\upd\ph'\intii\upd z' \inti \upd r'\,r'
      \abs{
      \colvec{0\\0\\1}
      \cdot\left[
        \colvec{\sqrt{1-\zb^2}\sin\phb\\
        \sqrt{1-\zb^2}\cos\phb\\ \zb}
        \times\colvec{r'\sin\ph'\\r'\cos\ph'\\z'}\right]}\\
      &\qquad\times\dirac{\rd-\sqrt{r'^2+z'^2}}\dirac{z'}
        \Heavi{\rd-|(r'\sin\ph',r'\cos\ph'-r,z'-z)|}\\
      &\qquad\times
      \dirac{\colvec{r'\sin\ph'\\r'\cos\ph'-r\\z'-z}\cdot
        \colvec{\sqrt{1-\zb^2}\sin\phb\\ \sqrt{1-\zb^2}\cos\phb \\ \zb}}
    \end{split}\label{eq:mayer:dd:simpleint}\\
    \begin{split}
    &=\frac{\rd}{2}\int_0^{2\pi}\upd\ph'\abs{
      \colvec{\sin\ph'\\ \cos\ph'\\ 0}
      \cdot\colvec{-\sqrt{1-\zb^2}\cos\phb\\ \sqrt{1-\zb^2}\sin\phb\\0}}\\
      &\qquad\times\Heavi{\rd-|(\rd\sin\ph',\rd\cos\ph'-r,-z)|}\\
      &\qquad\times\dirac{\colvec{\rd\sin\ph'\\ \rd\cos\ph'-r\\-z}
        \cdot\colvec{\sqrt{1-\zb^2}\sin\phb\\ \sqrt{1-\zb^2}\cos\phb\\
        \zb}} \end{split}\label{eq:mayer:dd:manyphi}\\
    \begin{split}
    &=\frac{\rd}{2}\int_0^{2\pi}\upd\ph'\abs{\sqrt{1-\zb^2}\sin(\phb-\ph')}
      \Heavi{2r\rd\cos\ph'-r^2-z^2}\\
      &\qquad\times
        \dirac{ r\sqrt{1-\zb^2}\cos\phb
          -\rd\sqrt{1-\zb^2}\cos(\ph'-\phb) +z\zb}
      \end{split}\label{eq:mayer:dd:singlephi}\\
    \begin{split}
    &=\frac{1}{2}\sum_\pm\Heavi{2r\rd\cos\left(\phb\pm\arccos
        \frac{r\sqrt{1-\zb^2}\cos\phb+z\zb}{\rd\sqrt{1-\zb^2}}\right)-r^2-z^2}\\
      &\qquad\times\Heavi{\rd\sqrt{1-\zb^2}-\abs{r\sqrt{1-\zb^2}\cos\phb+z\zb}},
    \end{split}\label{eq:dd:result}
\end{align}
where the integrations over $z'$ and $r'$ in
Eq.~(\ref{eq:mayer:dd:simpleint}) are straightforward.  From
(\ref{eq:mayer:dd:manyphi}) to (\ref{eq:mayer:dd:singlephi}) we
rewrote the argument of the Dirac delta distribution such that $\ph'$
appears only once, using the identity
$\sin\ph'\sin\phb+\cos\ph'\cos\phb=\cos(\ph'-\phb)$.

Given the complexity of the overlap condition between two arbitrarily
oriented platelets in three dimensions, it is not surprising that
Eq.~(\ref{eq:dd:result}) is an involved expression. It can be viewed
as counting the number of intersections between the platelet $D'$ and
the rim of platelet $D$
\footnote{We use $D$ and $D'$ to refer both to the platelets and the
points of their origins.  Consider the triangle $\Delta$ between the
center of platelet $D$, the intersection $I$ of platelet $D'$ with the
rim of $D$, and the center of $D'$ projected onto the $x$--$y$ plane
(call this point $A$).  This triangle has by construction two known
sides, $\rd$ and $r$.  The length of the third side of $\Delta$ can be
calculated by the cosine theorem from the angle $\angle ADI$.  Using
the Pythagorean theorem on the triangle between $I$ and the real and
projected centers of $D'$, the length $|\overline{D'I}|$ can be
calculated and compared to the radius of $D'$, $\rd$.  This is
accomplished by the first step function.  In order to calculate the
angle needed for the cosine theorem, consider $\Delta$ and the
right-angled triangle $\Delta'$ that is obtained by continuing
$\overline{AI}$ over $I$ to form a right angle at the new point $B$.
Then, $|\overline{AB}|=|r\cos\phb|$ and
$|\overline{AI}|=|z\zb|/\sqrt{1-\zb^2}$.  Therefore, the $\arccos$
equals the angle $\angle DIB$, and we obtain the angle $\angle ADI$ by
adding $\phb$. The second step function in Eq.~(\ref{eq:dd:result})
checks if the plane defined by $D'$ intersects with the rim of $D$.}.
The second term in Eq.\ (\ref{EQfDD}) counts the number of
intersections between platelet $D$ and the rim of platelet $D'$. Hence
in total there are two (zero) intersections for an overlapping
(nonoverlapping) configuration, and indeed the Mayer bond between
platelets, Eq.\ (\ref{EQfDD}), is recovered through convolution of
(orientation-dependent) single-particle measures.

\subsection{The platelet-needle Mayer bond}
\label{SECplateletNeedleMayerBond}
\begin{figure}
  \includegraphics[width=\imagewidth]{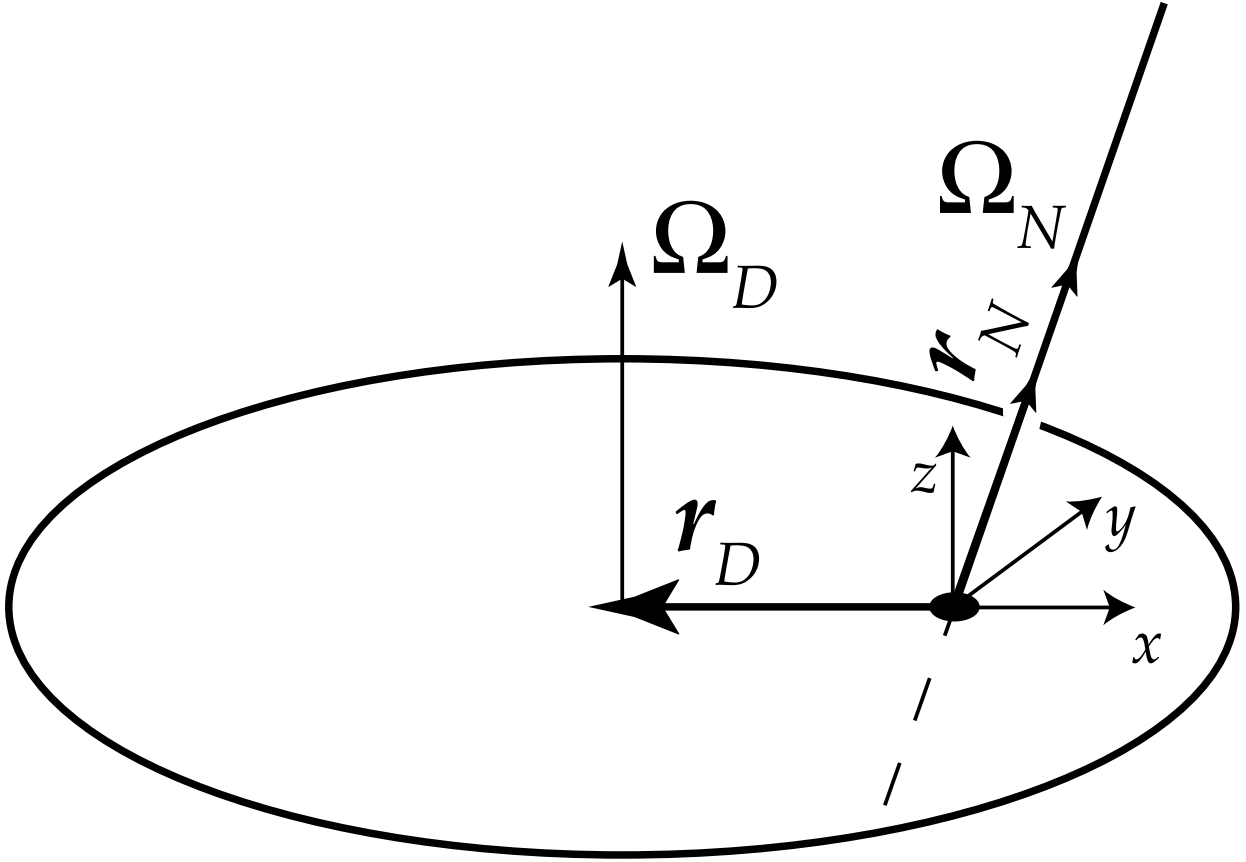}
  \caption{ Setup for the calculation of the platelet--needle Mayer
    bond.  The needle is located at $\rv_\needle$ in such a way that
    it points at the origin.  The platelet is located at $\rv_D$.
    }
  \label{fig:dn}
\end{figure}

We next consider the platelet-needle Mayer bond,
$f_{\diskneedle}(\rv,\Ov_D;\Ov_N)$, where $\rv$ is the difference
vector from disk center to needle center, and $\Ov_D$ and $\Ov_N$ are
the orientations of the disk and needle, respectively. We define a
``mixed'' weight function for the platelets that is non-vanishing on
the platelet surface, but carries an additional dependence on the rod
orientation,
\begin{equation}
  w_2^{\diskneedle}(\rv,\Od;\On) =
      2 |\Od\cdot\On|w_2^\disk(\rv,\Od),
  \label{EQw2DN}
\end{equation}
where $w_2^D(\rv,\Ov)$, as given in Eq.\ (\ref{EQw2DRosenfeld}),
describes the platelet surface. This allows us to obtain the Mayer
bond between platelet and needle via
\begin{equation}
  -f_{\diskneedle}(\rv,\Od;\On) 
    = w_2^{\diskneedle}(\rv,\Od;\On) \ast 
      w_1^\needle(\rv,\On).
  \label{EQfDN}
\end{equation}
The validity of Eq.\ (\ref{EQfDN}) can be seen by employing
cylindrical coordinates and putting the platelet on the $y$-axis, its
orientation being along the $z$-axis.  Position the needle in such a
way that its intersection with the $x$--$y$ plane is the origin (cf.\
Fig.~\ref{fig:dn}).
The needle is described by
$\rv_\needle=(z_\needle/\zb\sqrt{1-\zb^2}\sin\phb,
z_\needle/\zb\sqrt{1-\zb^2}\cos\phb,z_\needle)$ and
$\On=(\sqrt{1-\zb^2}\sin\phb, \sqrt{1-\zb^2}\cos\phb, \zb)$.  The
coordinates of the platelet are $\rv_D=(0,r_D,0)$ and
$\Od=(0,0,1)$; the integration variable is
$\xv=(x'\sin\ph',x'\cos\ph',z')$. From (\ref{EQw2DN}) and
(\ref{EQw2DRosenfeld}) we obtain the explicit expression
$w_2^{\diskneedle}(\rv,\Od;\On) =4\Heavi{\rd-\abs{\rv}}
\dirac{\rv\cdot\Od} \abs{\Od\cdot\On}$, that we insert together with
$w_1^N(\rv,\Ov_N)$, as given in Eq.\ (\ref{EQw1N}), into Eq.\ (\ref{EQfDN})
to obtain
\begin{align}
-f_{\diskneedle}&(\rv,\Od;\On)=
  \int \upd^3x\, w_1^\needle(\xv-\rv_\needle,\On)  
    w_2^{\diskneedle}(\xv-\rv_D,\Od;\On)\\
  \begin{split}
  &=\intii\upd z' \int_0^{2\pi}\upd\ph'\inti \upd x'\,x' 
    \int_{-L/2}^{L/2}\upd l\\
    &\quad\times
    \dirac{ x'\sin\ph' -z_\needle/\zb\sqrt{1-\zb^2}\sin\phb+l\sqrt{1-\zb^2}\sin\phb}\\
    &\quad\times
    \dirac{x'\cos\ph'-z_\needle/\zb\sqrt{1-\zb^2}\cos\phb+l\sqrt{1-\zb^2}\cos\phb}\\
    &\quad\times\delta(z'-z_\needle+l\zb)
    \Heavi{\rd-
      |(x'\sin\ph',x'\cos\ph'-r_D,z')|} |\zb|\delta(z')
  \end{split}\\
  \begin{split}\label{dn:mayer:midint}
  &=\int_0^{2\pi}\upd\ph'   
      \inti \upd x'\,x'\dirac{x'\sin\ph'}\dirac{x'\cos\ph'}\\
    &\quad \times \Heavi{\rd-|(x'\sin\ph',x'\cos\ph'-r_D)|}
      \Heavi{\frac{L}{2}|\zb|-|z_\needle|}
   \end{split}\\
  &=\Heavi{\rd-r_D}\Heavi{\frac{L}{2}|\zb|-|z_\needle|},
  \label{eq:dn:result}
\end{align}
where the integrals over $z'$ and $l$ are straightforward to
calculate.  In Eq.~\eqref{dn:mayer:midint}, the integral over $\ph'$
is performed first, rendering the $x'$-integral trivial.

The result, Eq.\ (\ref{eq:dn:result}), is indeed the Mayer bond
between platelet and needle: Due to the set-up, an overlap can occur
only at the origin and the step functions in Eq.\ (\ref{eq:dn:result})
provide a means to assess whether or not this is the case. The first
step function checks whether the origin is inside the platelet while
the second step function checks whether the needle intersects the
origin.

\subsection{The sphere-platelet Mayer bond}
\label{SECmayerSD}
\begin{figure}
  \includegraphics[width=\imagewidth]{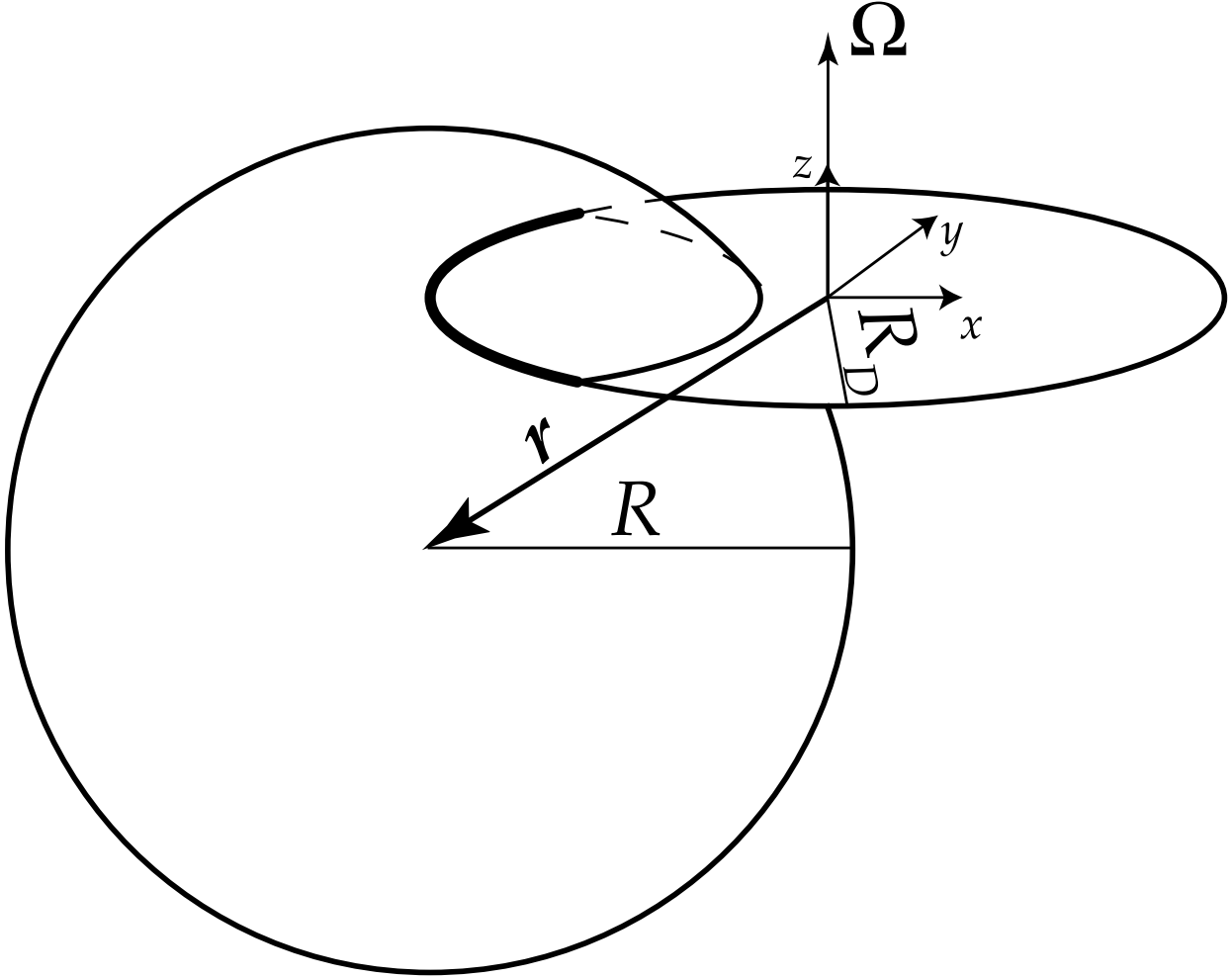}
  \caption{
      Setup for the calculation of the sphere--platelet Mayer bond.
      The platelet is located at the origin.
      The convolution $w_0^D*w_3^\sphere$ is equal to the angle under which
      the arc shown in bold appears.
      }
  \label{fig:sd:sphere}
\end{figure}

\begin{figure}
  \includegraphics[width=\imagewidth]{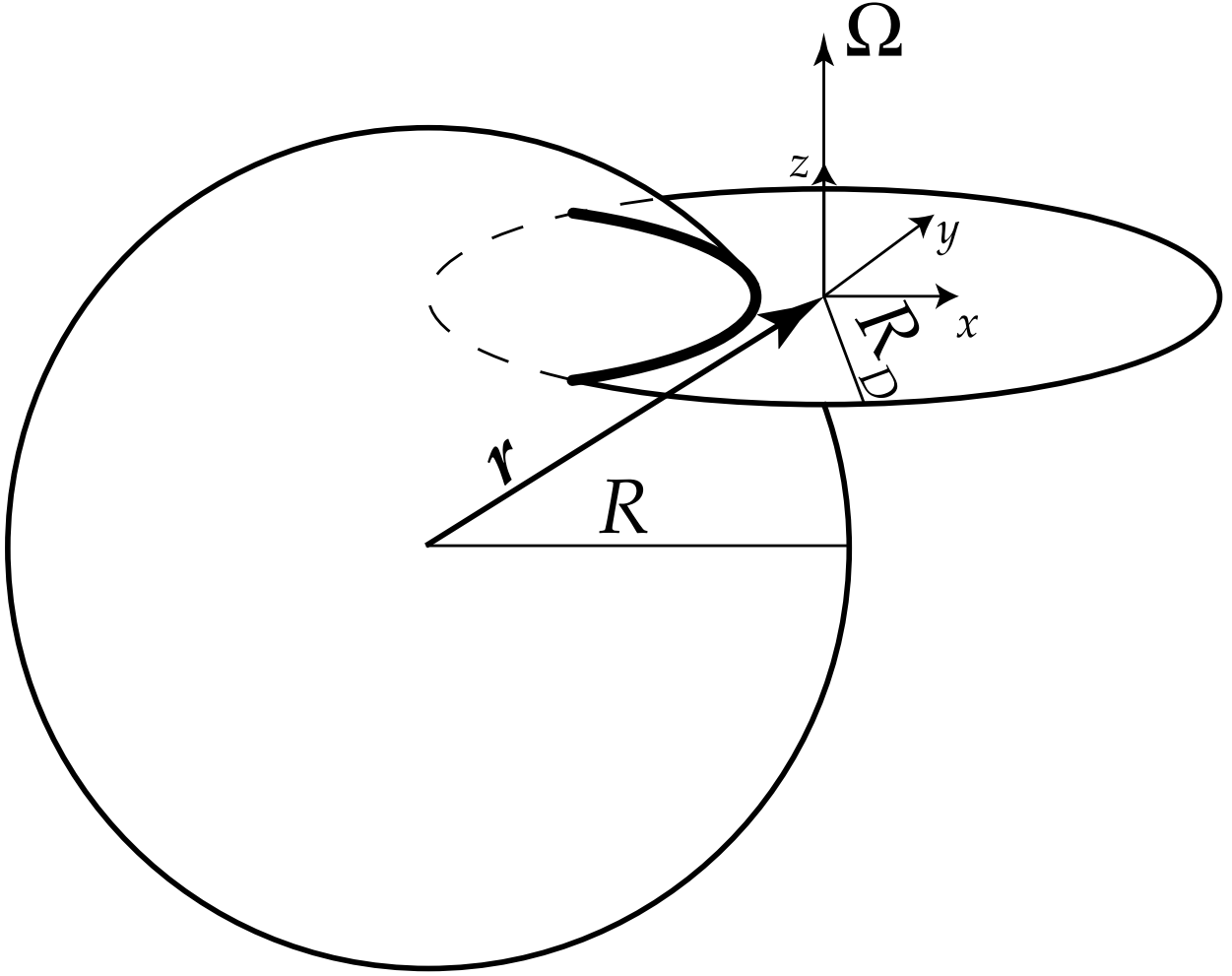}
  \caption{
      Setup for the calculation of the sphere--platelet Mayer bond.
      The platelet is located at the origin.
      The convolution $w_1^\sphere*w_2^D$ is equal to the angle under which
      the arc shown in bold appears.
      }
  \label{fig:sd:platelet}
\end{figure}

Treating the sphere-platelet Mayer bond leads to the most involved
geometry of pairwise particle overlap in the ternary mixture under
consideration.  It turns out that this case is intimately related to
the Mayer bond of the strictly two-dimensional hard disk mixtures.  As
detailed in Sec.\ \ref{SECtwod} the FMT gives only an approximate
representation in this case. Still the Rosenfeld functional for
two-dimensional hard disks is a reasonably accurate theory, see e.g.\
\cite{rasmussen02} for a recent study. Here we deal with this problem
on the same level of approximation as the two-dimensional Rosenfeld
case. Note that cutting the sphere with the plane of the platelet
yields a disk as the shape of intersection. We need to consider its
overlap with a platelet, and introduce the weight functions
\begin{align}
  w_2^{\spheredisk}(\rv,\Ov) &= 
  \frac{4}{\pi}
  \sqrt{1 - (\rv\cdot\Ov/R)^2} w_2^\sphere(\rv), 
   \label{EQw2SD}\\
  \wv{2}{\spheredisk}(\rv,\Ov) &= 
  4 \frac{\rv-(\rv\cdot\Ov)\Ov}{\pi R}w_2^\sphere(\rv),
   \label{EQwv2SD} \\
  \wv{1}{D}(\rv,\Ov) &= \frac{\rv}{\rd}w_1^D(\rv,\Ov),
   \label{EQwv1D}
\end{align}
where $w_2^S(\rv)$, as given in Eq.\ (\ref{EQw2S}), describes the
sphere surface, and $w_1^D(\rv,\Ov)$, as given in Eq.\
(\ref{EQw1DRosenfeld}), describes the platelet rim.  Eq.\
(\ref{EQwv2SD}) defines a vector field tied to the surface of the
sphere (with radius $R$). In contrast to the ``radial hedgehog'' of the
(classic) vector weight function $\wv{2}{\sphere}(\rv)$, as given in
Eq.\ (\ref{EQwv2S}), the direction of the current vector field is
radial with respect to the (platelet) $\Ov$ direction and its
magnitude decreases towards either pole -- a configuration one could
refer to as a ``cylindrical hedgehog''.  Note further that
$w_2^{\spheredisk}=|\wv{2}{\spheredisk}|$, in accordance with
$w_2^S=|\wv{2}{S}|$. The vector-field of Eq.\ (\ref{EQwv1D}) resembles
a corona and is the straightforward generalization of the
corresponding two-dimensional hard disk weight function,
$\wv{1}{(i)}(\rv)$, given in Eq.\ (\ref{EQwv1twod}). We use these
functions to approximate $f_{\spheredisk}\approx f_{\spheredisk}^\ast$
with
\begin{align}
 -f_{\spheredisk}^\ast(\rv,\Ov) & =
    w_3^\sphere(\rv) \ast w_0^D(\rv,\Ov) + w_1^\sphere(\rv) \ast w_2^D(\rv,\Ov)
    \nonumber\\ & \quad
    +  w_2^{\spheredisk}(\rv,\Ov) \ast w_1^D(\rv,\Ov) 
      - \wv{2}{\spheredisk}(\xv,\Ov) \ast
        \wv{1}{D}(\rv,\Ov),
 \label{EQfSDapproximation}
\end{align}
where again the convolution implies a scalar product between vectors.
The three terms on the r.h.s.\ of Eq.\ (\ref{EQfSDapproximation})
represent the arc that the rim of the platelet traces inside the
sphere, the arc that the sphere traces on the platelet (as illustrated
in Figs.~\ref{fig:sd:sphere} and \ref{fig:sd:platelet}, respectively),
and an additional contribution of the cusps where the two arcs meet.
We put the platelet into the $x$--$y$ plane with its center at the origin.
The 
center of the
sphere is located in the $y$--$z$ plane, see Fig~\ref{fig:sd:sphere};
the coordinates chosen are $\rv=(0,r,z)$, $\Ov=(0,0,1)$,
$\xv=(r'\sin\ph',r'\cos\ph',z')$.
We obtain for the first term on the r.h.s.\ of Eq.\ (\ref{EQfSDapproximation})
\begin{align}
  \int \upd^3x\,&w_0^D (\xv,\Ov) w_3^\sphere(\xv-\rv)\nonumber\\
    \begin{split}
    &=\frac{1}{2\pi\rd}\int_0^{2\pi}\!\upd\ph'\intii \! \upd z'\inti \!r'\,\upd r'
        \dirac{\rd-\sqrt{r'^2+z'^2}}
      \dirac{\xv\cdot\Ov}\\
      &\qquad\times\Heavi{R-\abs{\xv-\rv}}
    \end{split}\\
    &=\frac{1}{\pi}\arccos\Bigl(\frac{r^2+\rds{2}+z^2-R^2}{2r\rd}\Bigr)
        \Heavi{2r\rd-\abs{r^2+\rds{2}+z^2-R^2}}.
\end{align}
By the cosine theorem, this is equal (up to a factor of $2\pi$) to the
length of the arc which the rim of the disc traces inside the sphere,
see Eq.~\eqref{eq:2d:arc}. Note also that $R^2-z^2$ is the squared
radius of the circle that the platelet cuts out of the sphere.

For the second term on the r.h.s.\ of Eq.\ (\ref{EQfSDapproximation})
we obtain
\begin{align}
  \int \upd^3x\,&w_1^\sphere(\xv) w_2^D(\xv-\rv,\Ov)\nonumber\\
      \begin{split}
    &=\frac{1}{4\pi R}\intii \upd z'\int_0^{2\pi} \upd\ph'\inti \upd r'\,r'
        \dirac{R-\sqrt{r'^2+z'^2}}\\
      &\qquad\times
      2\Heavi{\rd-\sqrt{r^2+r'^2+(z'-z)^2-2rr'\cos\ph'}}
        \dirac{z'-z}
      \end{split}\\
    &=\frac{1}{2\pi}\int_0^{2\pi}\upd\ph'\Heavi{R-|z|}
        \Heavi{\rd-\sqrt{r^2+R^2-z^2-2r\sqrt{R^2-z^2}\cos\ph'}}\\
    &=\frac{1}{\pi}\arccos\Bigl(\frac{r^2+R^2-z^2-\rds{2}}{2r\sqrt{R^2-z^2}}\Bigr)
        \Heavi{2r\sqrt{R^2-z^2}-\abs{r^2+R^2-z^2-\rds{2}}},
\end{align}
which is indeed the length of the arc that the sphere traces on the
platelet divided by $2\pi$, see Eq~\eqref{eq:2d:arc}.

The relevant weight functions to treat the third 
and the fourth
term on the r.h.s.\
of Eq.\ (\ref{EQfSDapproximation}) are explicitly
$\wv{2}{\spheredisk}(\rv,\Ov)=4\dirac{R-|\rv|}[\rv-(\rv\cdot\Ov)\Ov]/(\pi
R)$ and
$\wv{1}{D}(\rv,\Ov)=\dirac{\rd-|\rv|}\dirac{\rv\cdot\Ov}\rv/(8\rd)$,
hence
\begin{align}
\int \upd^3x\, \bigl[&w_2^{\spheredisk}(\xv)w_1^D(\xv-\rv)
        -\wv{2}{\spheredisk}(\xv)\wv{1}{D}(\xv-\rv)\bigr]
        \nonumber\\
    \begin{split}
    &=\frac{1}{2\pi}
      \int_0^{2\pi}\upd\ph'\inti \upd r'\,r'\intii\upd z'
          \dirac{R-\sqrt{r'^2+z'^2}}\\
       &\qquad\times\dirac{\rd-\sqrt{r^2+r'^2-2rr'\cos\ph'+(z'-z)^2}}\\
       &\qquad\times\dirac{z'-z}
              \left(\frac{r'}{\sqrt{r'^2+z'^2}}
                -\frac{\colvec{r'\sin\ph'\\r'\cos\ph'\\0}
                    \cdot\colvec{r'\sin\ph'\\ r'\cos\ph'-r\\z'-z}} 
                  {\rd\sqrt{r'^2+z'^2}}\right)
    \end{split}\\
    \begin{split}
    &=\frac{1}{2\pi}
       \int_0^{2\pi}\upd\ph'\inti\,r'\,\upd r'
          \dirac{R-\sqrt{r'^2+z^2}}\\
       &\qquad\times\dirac{\rd-\sqrt{r^2+r'^2-2rr'\cos\ph'}}
              \left(\frac{r'}{\sqrt{r'^2+z^2}}-\frac{r'^2-rr'\cos\ph'} 
                  {\rd\sqrt{r'^2+z^2}}\right)
    \end{split}\\
    \begin{split}
    &=\frac{R}{2\pi}\int_0^{2\pi}\upd\ph'
       \dirac{\rd-\sqrt{r^2+R^2-z^2-2r\sqrt{R^2-z^2}\cos\ph'}}\\
       &\qquad\times
              \frac{\sqrt{R^2-z^2}}{R}
              \left(1-\frac{\sqrt{R^2-z^2}-r\cos\ph'}
             {\rd}\right)
             \Heavi{R-|z|}
    \end{split}\\
    \begin{split}
    &=\frac{1}{\pi}
   \Heavi{2r\sqrt{R^2-z^2}-|r^2+R^2-z^2-\rds{2}|}
        \Heavi{R-|z|}\\
       &\qquad\times\frac{R^2-(\sqrt{R^2-z^2}-\rd)^2}
                {\sqrt{4r^2(R^2-z^2)-(r^2+R^2-z^2-\rds{2})^2}}
    \end{split}\\
    &=\frac{1}{\pi} \Heavi{\rd-|\rds{\prime}-r|}
        \frac{r^2-(\rds{\prime}-\rd)^2}
          {\sqrt{4r^2\rds{\prime2}-(r^2+\rds{\prime2}-\rds{2})^2}}.
\end{align}
Considering again that $\rds{\prime2}=R^2-z^2$ is the squared radius
of the circle that the plane of the platelet cuts out of the sphere,
we recover Rosenfeld's approximation for the Mayer bond of hard disks
in two dimensions, see Eq.~\eqref{eq:2d:cusp}. Hence the same defect,
as displayed in Fig.\ \ref{FIG2d} for two-dimensional hard disks, is
encountered for the overlap between sphere and platelet. Based on the
very reasonable quality of the two-dimensional Rosenfeld functional,
we expect this to do little harm in typical applications of the
theory.

\section{Third virial level: Triangle diagrams}	
\label{SECtriangle}
The contribution of third order in densities to the excess free energy
of a mixture has the generic exact form of a sum of triangle diagrams,
\begin{equation}
 -\frac{1}{6} \sum_{ijk}
 \int \upd \sv \rho_i(\sv) \int \upd \sv' \rho_j(\sv')
 \int \upd \sv'' \rho_k(\sv'')
 f_{ij}(\sv,\sv') f_{jk}(\sv',\sv'') f_{ik}(\sv,\sv''),
 \label{EQfexcThirdVirialMixture}
\end{equation}
where $i,j,k$ label the species, the sum runs over all species,
$f_{ij}(\sv,\sv')$ is the Mayer bond between particles of species $i$
and $j$ with coordinates $\sv$ and $\sv'$, respectively, and $\sv$
(and its primed versions) is a composite variable appropriate for the
species under consideration, i.e.\ for the case of (uniaxial) rotators
like platelets $\sv=(\rv,\Ov)$, $\int \upd\sv = \int \upd\rv \int \upd
\Ov/(4\pi)$, and for spheres simply $\sv=\rv$, $\int \upd \sv = \int
\upd \rv$.

Even for one-component hard spheres FMT yields only an approximation
to (\ref{EQfexcThirdVirialMixture}) due to the occurrence of ``lost
cases'' \cite{tarazona97,cuesta02}: The FMT approximation for the
triple product of Mayer bonds in (\ref{EQfexcThirdVirialMixture}) is
non-vanishing only for configurations $\rv$, $\rv'$, and $\rv''$, such
that three spheres with (hard sphere) radius $R$ centered at these
positions share a non-vanishing common volume of intersection. Note
that the existence of such triple intersection implies pairwise
intersections, but not vice versa: An illustrative example is that
constituted by three touching spheres; see e.g.\ \cite{cuesta02} for
an in-depth discussion. The lost cases are compensated by overcounting
the cases with triple intersection, such that a reasonable (exact in
the pure hard sphere case) third virial coefficient for the bulk fluid
results. The upshot is that new weight functions result, that can be,
via weighted densities, incorporated to improve the accuracy of the
density functional approximation.

Here we follow a similar strategy and aim at controlling the
dependence on positions {\em and} orientations through appropriately
constructed weight functions. For triplets of platelets in
configurations with common triple intersection we can express
\begin{align}
f_{DD}(\rv-\rv',&\Ov;\Ov')
f_{DD}(\rv-\rv'',\Ov;\Ov'')
f_{DD}(\rv'-\rv'',\Ov',\Ov'') \nonumber\\
& = A_{DDD} \int \upd \xv
 w_2^{DDD}(\rv-\xv,\Ov;\Ov';\Ov'') 
 w_2^D(\rv'-\xv,\Ov')
 w_2^D(\rv''-\xv,\Ov''),
 \label{EQtripleDDD}
\end{align}
where $A_{DDD}$ is a constant, $w_2^D(\rv,\Ov)$ is the surface weight
function defined in Eq.\ (\ref{EQw2DRosenfeld}), and we introduce an
additional surface weight that carries a dependence on {\em two}
further (platelet) orientations,
\begin{equation}
 w_2^{DDD}(\rv,\Ov;\Ov';\Ov'') = \frac{8}{\pi}
 |\Ov\cdot(\Ov'\times\Ov'')| w_2^D(\rv,\Ov),
 \label{EQw2DDD}
\end{equation}
with normalization such that $\int \upd \rv \int \frac{d\Ov}{4\pi}
\int \frac{d\Ov'}{4\pi} \int \frac{d\Ov''}{4\pi}
w_2^{DDD}(\rv,\Ov;\Ov';\Ov'')=\xi_2^D=2\pi R_D^2$.  Setting $A_{DDD} =
\pi/64 =0.0491$ yields equality in Eq.\ (\ref{EQtripleDDD}) for
particle configurations with common triple intersection of the three
particles shapes.  Considering the bulk equation of state for pure
platelets from SPT, and in particular its (approximative) third virial
coefficient, the compensation for the lost cases (those without common
triple intersection, but with pairwise intersections) results in
$A_{DDD}=\pi/32=0.0982$.  In the following we will rather use the
value from FMT, $A_{DDD}=1/(4\pi)=0.0796$, which is numerically
similar to the SPT result. The advantage of the FMT scheme is that it
consistently applies to mixtures, see Ref.\
\cite{oversteegen05cookbook} for a detailed discussion.

For the platelet-sphere mixture the triangle diagram stemming from a
pair of platelets and a single sphere, provided that there is a a
common intersection of all three particle shapes, can be expressed as
\begin{align}
&  f_{SD}(\rv;\rv',\Ov')f_{DD}(\rv',\Ov';\rv'',\Ov'')f_{SD}(\rv;\rv'',\Ov'')
  \nonumber\\ 
& = \int \upd \xv \Bigl[ A_{SDD}w_2^{SDD}(\rv-\xv;\Ov';\Ov'')
  w_2^D(\rv'-\xv,\Ov') w_2^D(\rv''-\xv,\Ov'')\nonumber\\
& \qquad\qquad + w_3^S(\rv-\xv) \Bigl(
  w_1^{DD}(\rv'-\xv,\Ov';\Ov'') w_2^D(\rv''-\xv,\Ov'')\nonumber\\
& \qquad\qquad\qquad +
  w_2^{D}(\rv'-\xv,\Ov') w_1^{DD}(\rv''-\xv,\Ov'',\Ov') \Bigr) \Bigr],
  \label{EQtripleSDD}
\end{align}
where $A_{SDD}$ is a constant, and we have introduced a sphere weight
function that depends on two platelet orientations,
\begin{equation}
 w_2^{SDD}(\rv;\Ov';\Ov'') = \frac{8}{\pi}
 |(\Ov'\times\Ov'')\cdot \wv{2}{S}(\rv)|,
 \label{EQw2SDD}
\end{equation}
with the normalization $\int \upd \rv \int \frac{d\Ov}{4\pi} \int
\frac{d\Ov'}{4\pi} w_2^{SDD}(\rv;\Ov';\Ov'')=\xi_2^S = 4\pi R^2$.
Again $A_{SDD}=\pi/64=0.0491$ yields equality in Eq.\
(\ref{EQtripleSDD}) provided that there is common triple intersection.
We will use the FMT compensation, $A_{SDD}=1/(4\pi)=0.0796$.

The remaining case is that of triplets consisting of a single platelet
and a pair of spheres. This case involves similar geometry as the
platelet-sphere overlap, and hence we do not aim at an exact
treatment. Again in the spirit of the two-dimensional hard disk case
(cf.\ Sec.\ \ref{SECtwod}), we use an approximation,
\begin{align}
&f_{SS}(\rv'-\rv'')f_{SD}(\rv';\rv,\Ov)f_{SD}(\rv'';\rv,\Ov)\nonumber\\
&\approx\int \upd \xv \Bigl[
w_3^S(\rv'-\xv) w_3^S(\rv''-\xv) w_0^D(\rv-\xv,\Ov)
\nonumber\\
& \qquad + w_3^S(\rv'-\xv) \Bigl(
 w_2^{SD}(\rv''-\xv;\Ov) w_1^D(\rv-\xv,\Ov) -
 \wv{2}{SD}(\rv''-\xv;\Ov) \cdot \wv{1}{D}(\rv-\xv,\Ov) \Bigr)
\nonumber\\
& \qquad + w_3^S(\rv''-\xv) \Bigl(
 w_2^{SD}(\rv'-\xv;\Ov) w_1^D(\rv-\xv,\Ov) -
 \wv{2}{SD}(\rv'-\xv;\Ov) \cdot \wv{1}{D}(\rv-\xv,\Ov) \Bigr)
\nonumber\\
& \qquad + A_{SSD}\; w_2^D(\rv-\xv,\Ov) \Bigl(
  w_2^{SD}(\rv'-\xv;\Ov) w_2^{SD}(\rv''-\xv,\Ov) 
\nonumber\\ 
&\qquad\qquad -\wv{2}{SD}(\rv''-\xv;\Ov)\cdot\wv{2}{SD}(\rv''-\xv;\Ov)
\Bigr)\Bigr],
\end{align}
where 
$A_{SSD}=1/16$ generates the best possible approximation and $A_{SSD}=1/(4\pi)$ 
compensates for the lost cases in the FMT approximation.

\section{Weighted densities}
\label{SECweightedDensities}
We first summarize the novel weight functions that describe the
platelet degrees of freedom in the ternary mixture. From the original
Rosenfeld prescription we have obtained [see Eqs.\
(\ref{EQw2DRosenfeld})-(\ref{EQw0DRosenfeld}) in Sec.\
\ref{SECplateletPlateletMayerBond}]
\begin{align}
  w_2^D(\rv,\Ov) &= 2\Theta(\rd-|\rv|)\delta(\rv\cdot\Ov),
  \label{EQSUMw2DRosenfeld}\\
  w_1^D(\rv,\Ov) &= \delta(\rd-|\rv|)\delta(\rv\cdot\Ov)/8,
  \label{EQSUMw1DRosenfeld}\\
  w_0^D(\rv,\Ov) &= \frac{1}{2\pi \rd}\delta(\rd-|\rv|)\delta(\rv\cdot\Ov).
  \label{EQSUMw0DRosenfeld}
\end{align}
In order to treat platelet-platelet and platelet-needle overlap we
have introduced [see Eq.\ (\ref{EQw1DD}) in Sec.\
\ref{SECplateletPlateletMayerBond} and Eq.\ (\ref{EQw2DN}) in Sec.\
\ref{SECplateletNeedleMayerBond}]
\begin{align}
  w_1^{DD}(\rv,\Ov;\Ov') = & 
    \frac{2}{\rd}|\Ov \cdot (\Ov'\times \rv)| w_1^D(\rv,\Ov),
    \label{EQSUMw1DD}\\
  w_2^{\diskneedle}(\rv,\Od;\On) &=
    2 |\Od\cdot\On|w_2^\disk(\rv,\Od),
    \label{EQSUMw2DN}
\end{align}
respectively. To treat sphere-platelet overlap we use [see Eq.\
(\ref{EQw2SD})-(\ref{EQwv1D}) in Sec.\ \ref{SECmayerSD}]
\begin{align}
  w_2^{\spheredisk}(\rv,\Ov) &= 
    \frac{4}{\pi}
    \sqrt{1 - (\rv\cdot\Ov/R)^2} w_2^\sphere(\rv), 
    \label{EQSUMw2SD}\\
  \wv{2}{\spheredisk}(\rv,\Ov) &= 
    4 \frac{\rv-(\rv\cdot\Ov)\Ov}{\pi R}w_2^\sphere(\rv),
    \label{EQSUMwv2SD}\\
  \wv{1}{D}(\rv,\Ov) &= \frac{\rv}{\rd}w_1^D(\rv,\Ov),
    \label{EQSUMwv1D}
\end{align}
The consideration of the third order contributions to the excess free
energy has led to [see Eqs.\ (\ref{EQw2DDD}) and (\ref{EQw2SDD})
in Sec.\ \ref{SECtriangle}]
\begin{align}
  w_2^{DDD}(\rv,\Ov;\Ov';\Ov'') &= \frac{8}{\pi}
    |\Ov\cdot(\Ov'\times\Ov'')| w_2^D(\rv,\Ov),
    \label{EQSUMw2DDD}\\
  w_2^{SDD}(\rv;\Ov';\Ov'') &= \frac{8}{\pi}
    |(\Ov'\times\Ov'')\cdot \wv{2}{S}(\rv)|.
    \label{EQSUMw2SDD}
\end{align}

All weight functions are used to build weighted densities via
convolutions with the bare density profiles.  For hard spheres the
scalar and vectorial weighted densities are respectively given by
\begin{align}
  n_\nu^\sphere(\rv) &= w_\nu^\sphere(\rv) \ast \rhos(\rv), 
  \quad \nu=0,1,2,3,\\
  {\bf n}_\nu^S(\rv) &= {\bf w}_\nu^S(\rv) \ast \rhos(\rv),
  \quad \nu=v1,v2,
\end{align}
where $w_\nu^\sphere(\rv)$ and ${\bf w}_\nu^S(\rv)$ as given in and
below Eqs.\ (\ref{EQw3S})-(\ref{EQwv2S}) are the classic Rosenfeld
hard sphere functions. Adding non-spherical particles requires
convolving the sphere density profile with orientation-dependent weight
functions. The coupling to needles is accomplished via
\begin{align}
  n_2^{\sphereneedle}(\rv,\Ov) &= w_2^{\sphereneedle}(\rv,\Ov) \ast \rhos(\rv),
\end{align}
where $w_2^{\sphereneedle}(\rv,\Ov)$ as given in Eq.\ (\ref{EQw2SN})
is a mixed sphere-needle function. Treating the platelets follows
a similar outline,
\begin{align}
  n_2^{SD}(\rv,\Ov) &= w_2^{SD}(\rv,\Ov) \ast \rhos(\rv),\\
  \nv{2}{SD}(\rv,\Ov) &= \wv{2}{SD}(\rv,\Ov) \ast \rhos(\rv),\\
  n_2^\SDD(\rv;\Ov;\Ov') &= w_2^\SDD(\rv;\Ov;\Ov') \ast \rho_\sphere(\rv),
\end{align}
with $w_2^{SD}(\rv,\Ov)$ and $\wv{2}{SD}(\rv,\Ov)$ given in Eqs.\
(\ref{EQw2SD}) and (\ref{EQwv2SD}), respectively, and
$w_2^\SDD(\rv;\Ov';\Ov'')$ as given in Eq.\ (\ref{EQw2SDD}).

For needles the pure weight functions are
\begin{align}
  n_\nu^N(\rv,\Ov) &= w_\nu^N(\rv,\Ov) \ast \rhon(\rv,\Ov), \quad
  \nu=0,1,
\end{align}
where $w_0^N(\rv,\Ov)$ and $w_1^N(\rv,\Ov)$ are given in Eqs.\
(\ref{EQw0N}) and (\ref{EQw1N}), respectively. In case one considers
needle-needle interactions, additional weighted densities need to be
taken into account \cite{brader02rsa}.

The degrees of freedom of the platelets are modelled through the
weighted densities
\begin{align}
  n_\nu^D(\rv,\Ov) &= w_\nu^D(\rv,\Ov) \ast \rhod(\rv,\Ov), 
  \quad \nu=0,1,2, \label{EQnnuD}\\
  \nv{1}{D}(\rv,\Ov) &= \wv{1}{D}(\rv,\Ov)\ast \rhod(\rv,\Ov),\\
  n_1^{DD}(\rv,\Ov') &= 
  \int\frac{\upd\Ov}{4\pi} w_1^{DD}(\rv,\Ov;\Ov') 
      \ast \rhod(\rv,\Ov),\label{EQn1DD}\\
  n_2^\DDD(\rv;\Ov;\Ov') &=  \int \frac{\upd\Ov''}{4\pi} 
    w_2^\DDD(\rv,\Ov'';\Ov;\Ov') \ast \rho_D(\rv,\Ov''),\label{EQn2DDD}
\end{align}
with the weight functions $w_2^D(\rv,\Ov)$, $w_1^D(\rv,\Ov)$, and
$w_0^D(\rv,\Ov)$, as given in Eqs.\ (\ref{EQw2DRosenfeld}),
(\ref{EQw1DRosenfeld}), and (\ref{EQw0DRosenfeld}), respectively, and
$\wv{1}{D}(\rv,\Ov)$ as given in Eq.\ (\ref{EQwv1D}).  The double
orientation-dependent weight function $w_1^{DD}(\rv,\Ov;\Ov')$ is
given in Eq.\ (\ref{EQw1DD}).  The weight function that we have
introduced to control the triangle diagrams,
$w_2^\DDD(\rv,\Ov';\Ov'';\Ov''')$, is given in Eq.\ (\ref{EQw2DDD}).

The remaining coupling of the platelets to the needles is achieved via
\begin{align}
  n_2^{DN}(\rv,\Ov') &= \int \frac{\upd\Ov}{4\pi}
    w_2^{DN}(\rv,\Ov;\Ov') \ast \rhod(\rv,\Ov),
\end{align}
where we have used $w_2^{DN}(\rv,\Ov)$ as given in Eq.\ (\ref{EQw2DN})
to build a mixed weighted density.

\section{Free energy functional}
\label{SECdft}
In Ref.\ \cite{schmidt01rsf} the excess Helmholtz free energy
functional for particles with rotational degrees of freedom was
expressed not only as a spatial integral (as was proposed in Refs.\
\cite{rosenfeld94convex,rosenfeld95convex}), but also as an integral
over the director space. In our present explicit consideration of
three particle correlations an additional integration over director
space is required,
\begin{equation}
  \beta F_{\rm exc} = \int\upd\rv\int\frac{\upd\Ov}{4\pi}
  \int\frac{\upd\Ov'}{4\pi} \Phi(\{n_\nu^i\})
  \label{EQfexcGeneric}
\end{equation}
where the (reduced) free energy density, $\Phi$, is a function of the
set of weighted densities, $\{n_\nu^i\}$, where $i$ labels the species
and $\nu$ labels the type of weight function.

For hard spheres $\Phi=\phifmt{\sphere}$ with
\begin{align}
  \phifmt{\sphere} = & -n_0^\sphere \ln(1-n_3^\sphere) + 
  \frac{n_1^\sphere n_2^\sphere-\nv{1}{\sphere}\cdot 
        \nv{2}{\sphere}}{1-n_3^\sphere}\nonumber\\ & + 
  \frac{\frac{1}{3}(n_2^\sphere)^3-n_2^\sphere 
      \nv{2}{\sphere}\cdot\nv{2}{\sphere}}{8\pi(1-n_3^\sphere)^2},
  \label{EQphiS}
\end{align}
where we have suppressed here and in the following the dependence of
the weighted densities on the space coordinate $\rv$.  Eq.\
(\ref{EQphiS}) is the original Rosenfeld form \cite{rosenfeld89} that
yields, for bulk fluids, the (reduced) free energy density as obtained
by the Percus-Yevick compressibility (or scaled-particle)
route. Improved versions exist that feature exact dimensional
crossover \cite{RSLTlong,tarazona00}, and the Carnahan-Starling
equation for bulk fluid states \cite{roth02whitebear}.

For binary mixtures of spheres and needles the reduced free energy
density is $\Phi=\phifmt{\sphere}+\phifmt{\sphereneedle}$,
where $\phifmt{\sphere}$ is given in
(\ref{EQphiS}), and $\phifmt{\sphereneedle}$ describes the effect 
of needle-sphere interactions \cite{schmidt01rsf}, given as 
\begin{equation}
  \phifmt{\sphereneedle} = -n_0^\needle(\Ov) \ln(1-n_3^\sphere) 
      + \frac{n_1^\needle(\Ov) n_2^{\sphereneedle}(\Ov)}{1-n_3^\sphere},
  \label{EQphiSN}
\end{equation}
where we suppress the dependence of the weighted densities on $\rv$,
but make the dependence on orientation explicit in the notation. The
free energy for bulk fluid states obtained from (\ref{EQphiSN}) is the
same as that obtained from a thermodynamic (free-volume-like)
perturbation theory \cite{bolhuis94}, that results in a fluid-fluid
demixing binodal that compares well with results from simulations.

For a system of pure platelets $\Phi=\phifmt{D}$, with
\begin{equation}
 \phifmt{D} = n_1^{DD}(\Ov)n_2^D(\Ov) + \frac{1}{24\pi}
  n_2^D(\Ov) n_2^{DDD}(\Ov,\Ov') n_2^D(\Ov'),
 \label{EQphiD}
\end{equation}
where the prefactor of the third order term in density is according to
Rosenfeld $A=1/(24\pi)=0.01326$.  Scaled-particle theory
\cite{barker76,oversteegen05cookbook} gives a slightly different
value, namely $A=\pi/192=0.01636$; see Ref.\
\cite{oversteegen05cookbook} for a detailed comparison of the
predictions of these approaches for bulk fluid properties. Eq.\
(\ref{EQphiD}) is the simplest form of a free energy density that
features the correct second virial level and a similar free energy for
bulk isotropic fluid states as scaled-particle theory.

For binary sphere-platelet mixtures we obtain $\Phi=\phifmt{\sphere} +
\phifmt{\spheredisk}$, where the hard sphere contribution $\Phi_S$ is
given in Eq.\ (\ref{EQphiS}), and
\begin{align}
 \phifmt{\spheredisk} =& -n_0^D(\Ov) \ln(1-n_3^\sphere) \nonumber \\ &
   + \frac{n_1^\sphere n_2^D(\Ov)
   + n_2^{\spheredisk}(\Ov) n_1^D(\Ov)
   - \nv{2}{\spheredisk}(\Ov)\cdot\nv{1}{D}(\Ov)}{1-n_3^\sphere} 
   + \frac{n_1^{DD}(\Ov)n_2^D(\Ov)}{1-n_3^\sphere}
   \nonumber \\ &
   + \frac{\Bigl(
     n_2^\SD(\Ov) n_2^\SD(\Ov) - \nv{2}{\SD}(\Ov) \cdot \nv{2}{\SD}(\Ov)
     \Bigr) n_2^D(\Ov)}{8\pi(1-n_3^\sphere)^2}
   \nonumber \\ &  
   + \frac{n_2^\SDD(\Ov,\Ov') n_2^D(\Ov) n_2^D(\Ov')}
      {8\pi(1-n_3^\sphere)^2}
   + \frac{n_2^\DDD(\Ov,\Ov') n_2^D(\Ov) n_2^D(\Ov')}
      {24\pi(1-n_3^\sphere)^2}.
  \label{EQphiSD}
\end{align}

For a binary platelet--needle mixture the free energy density is
$\Phi=\Phi_D+\Phi_{\diskneedle}$, where the pure platelet
contribution $\Phi_D$ is given in Eq.\ (\ref{EQphiD}), and a simple
second-order coupling between both species is introduced via
\begin{align}
  \phifmt{\diskneedle}=n_2^{\diskneedle}(\Ov;\Ov')n_1^\needle(\Ov').
\end{align}
Due to the vanishing thickness of particles of both species, any
higher order terms vanish according to SPT.

Finally, for ternary sphere-platelet-needle mixtures the free energy
density is
\begin{equation}
  \Phi = \phifmt{\sphere} + \phifmt{\sphereneedle} 
      + \phifmt{\spheredisk} 
      + \frac{n_2^{\diskneedle}(\Ov;\Ov') n_1^\needle(\Ov')}{1-n_3^\sphere},
\end{equation}
where the first three terms on the r.h.s.\ are given through Eqs.\
(\ref{EQphiS}), (\ref{EQphiSN}) and (\ref{EQphiSD}), respectively.

\section{Planar geometry}
\label{SECplanarGeometry}
In many practical situations one is faced with inhomogeneities that
depend only on a single (Cartesian) coordinate, while being
translationally invariant in the two remaining directions. A smooth
planar wall, where the fluid density profile(s) only depend on the
perpendicular distance, $z$, from the wall is a primary example. For
rotators an additional simplification arises from (cylindrical)
rotational symmetry around the $z$-axis.  Hence such problems are
characterized solely by $z$, and the tilt angle $\vartheta$ between
the particle orientation and the $z$-axis.  We give in the following
explicit expressions for (reduced) weight functions appropriate for
efficient numerical treatment of such situations.  For completeness we
also give the results for spheres and needles \cite{brader02rsa}.

Let us start by defining the full set of reduced densities; we use the
parametrizations $\rv=(x,y,z)$ and $\Ov=(\vartheta,\varphi)$. For
the spheres
\begin{align}
 w_\nu^S(z) = \int \upd x \int \upd y \, w_\nu^S(\rv), \quad
 \nu = 3,2,1,0,v2,v1.
\end{align}
Further mixed weight functions couple spheres to needles,
\begin{align}
  w_2^{SN}(z,\vartheta) &= \int\upd x\int\upd y\, w_2^{SN}(\rv,\Ov),
\end{align}
and spheres to platelets
\begin{align}
  w_2^{SD}(z,\vartheta) &= \int\upd x\int\upd y\,w_2^{SD}(\rv,\Ov),\\
  \wv{2}{SD}(z,\vartheta) &= \int\frac{\upd \ph}{2\pi}
  \int\upd x\int \upd y\, \wv{2}{\spheredisk}(\rv,\Ov),\\
  w_2^{SDD}(z,\vartheta,\vartheta') &=
  \int\frac{\upd\ph}{2\pi}\int\frac{\upd\ph'}{2\pi}\int\upd x\int\upd
  y\,w_2^{SDD}(\rv,\Ov,\Ovv),
\end{align}

For the needles
\begin{align}
  w_\nu^N(z,\vartheta) &= \int \upd x\int \upd y\, w_\nu^N(\rv,\Ov), 
  \qquad \nu=0,1,
\end{align}

For the platelets the effective weight functions for planar geometry
are obtained by integrating over the lateral coordinates
\begin{equation}
  w_\nu^D(z,\vartheta) = \int\upd x\int \upd y\,w_\nu^D(\rv,\Ov),
    \qquad \nu=0,1,2,
  \label{EQwDiskEffective}
\end{equation}
and
\begin{align}
  \wv{1}{D}(z,\vartheta) &= \int \frac{\upd \ph}{2\pi} \int\upd x\int \upd y\,
      \wv{1}{D}(\rv,\Ov),\\
  w_1^{DD}(z,\vartheta,\vartheta') &= \int\frac{\upd\ph'}{2\pi}\int\upd
  x\int\upd y\, w_1^{DD}(\rv,\Ov,\Ov')\\
  w_2^{DDD}(z,\vartheta,\vartheta',\vartheta'') &=
    \int\frac{\upd\ph'}{2\pi}\int\frac{\upd\ph''}{2\pi}
      \int\upd x\int\upd y\, w_2^{DDD}(\rv,\Ov,\Ov',\Ov'')
\end{align}

\begin{align}
  w_2^{DN}(z,\vartheta_D; \vartheta_\needle) &=
      \int\frac{\upd \ph_D }{2\pi} \int\upd x\int \upd y\,
        w_2^{DN}(\rv,\Ov_D; \Ov_\needle).
\end{align}

Using the effective weight functions the weighted densities can then
be written as
\begin{align}\label{eq:planar:nvSD}
  \nv{2}{SD}(z,\vartheta) &=  \wv{2}{SD}(z,\vartheta)\ast\rhos(z),
\end{align}

\begin{align}
  n_\nu^D(z,\vartheta) &=  w_\nu^D(z,\vartheta)  \ast \rhod(z,\vartheta), 
   \quad \nu=0,1,2,\\
  \nv{1}{D}(z,\vartheta) &=  \wv{1}{D}(z,\vartheta) \ast \rhod(z,\vartheta),\\
  n_1^{DD}(z,\vartheta') &= 
  \frac{1}{2}\int_0^\pi \upd\vartheta\, \sin\vartheta\,
  w_1^{DD}(z,\vartheta;\vartheta') \ast \rhod(z,\vartheta),\\
  n_2^{DDD}(z,\vartheta,\vartheta')
  &=\frac{1}{2}  \int_0^\pi\upd\vartheta''\,
    w_2^{DDD}(z,\vartheta'';\vartheta;\vartheta')
      \ast\rho_D(z,\vartheta'')\\
  n_2^{DN}(z,\vartheta') &= 
    \frac{1}{2}\int_0^\pi \upd\vartheta\, \sin\vartheta\,
     w_2^{DN}(z,\vartheta;\vartheta') \ast \rhod(z,\vartheta),
     \label{eq:planar:nDN}
\end{align}
where in Eqs.~\eqref{eq:planar:nvSD}-\eqref{eq:planar:nDN} (and only
here) $\ast$ denotes the one-dimensional convolution, $g(z)\ast
h(z)=\int g(z'-z)h(z')$. In contrast to the needle case the
integrands depend non-trivially on $\varphi$.

We finally give all results for the effective weight functions in
planar geometry. For hard spheres the well-known results are
\begin{align}
 w_3^S(z) &= \pi(R^2-z^2)\Theta(R-|z|),\label{EQw3sPlanar}\\
 w_2^S(z) &= 2\pi R\Theta(R-|z|),\label{EQw2sPlanar}\\
 \wv{2}{\sphere}(z) &= 2\pi z\Theta(R-|z|){\bf e}_z,\label{EQwv2Planar}
\end{align}
where ${\bf e}_z$ is the unit vector pointing along the $z$-axis.  The
linearly dependent weight functions are $w_1^S(z)=\Theta(R-|z|)/2$,
$w_0^S(z)=\Theta(R-|z|)/(2R)$, $\wv{1}{\sphere}(z)=z\Theta(R-|z|){\bf
e}_z/(2R)$.  The mixed sphere-needle weight function is given through
\begin{equation}
w_2^{SN}(z,\vartheta) =\begin{cases}
8\sqrt{R^2\sin^2\vartheta-z^2} &\\ \quad +8z\cos\vartheta&\\
\qquad\times\arcsin\Bigl(\frac{z\cot(\vartheta)}{\sqrt{R^2-z^2}}\Bigr)
& \text{if }\- |z| < R\sin\vartheta\\
4\pi|z|\cos\vartheta & \text{if } R\sin\vartheta\leq|z|\leq R\\
0 & \text{otherwise.}
\end{cases}
\end{equation}
Further mixed sphere weight functions are
\begin{align}
  w_2^{SD}(z,\vartheta)
  &=\frac{2}{\pi^2}\Theta(R-|z|) \int_0^{2\pi} d\varphi
  \sqrt{R^2-\Bigl[(R^2-z^2) \cos \varphi \sin \vartheta + 
      z \cos \vartheta\Bigr]^2},\\
  \wv{2}{SD}(z,\vartheta)
      &=\frac{8\pi}{\rd}z\sin^2\vartheta\sqrt{\rds{2}-z^2}\Theta(R-|z|)\ev_z,\\
  w_2^{SDD}(z,\vartheta,\vartheta') &=
  \Heavi{R-|z|}\int_0^{2\pi}\upd\ph''
    \begin{cases}
      \frac{8}{\pi}|t| &\text{if }t^2> s^2\\
      \frac{16}{\pi^2}\Bigl(\sqrt{s^2-t^2} 
          +t \arcsin\frac{t}{s}\Bigr)
        &\text{else,}
    \end{cases}
\end{align}
where $\ev_z$ is a unit vector in the $z$-direction,
$t=z\sin\vartheta\sin\vartheta'\sin\ph'$, and
$s=\sqrt{R^2-z^2}\sqrt{\sin^2\vartheta\cos^2\vartheta'
  +\cos^2\vartheta\sin^2\vartheta'
  -2\sin\vartheta\cos\vartheta\sin\vartheta'\cos\vartheta'\cos\ph'}$.

For the needles the effective weight functions are
\begin{align}
w_1^N(z,\vartheta) &= (4\cos\vartheta)^{-1}
 \Theta\Bigl(\frac{L}{2}\cos\vartheta-|z|\Bigr), \label{EQw1NPlanar}\\
w_0^N(z,\vartheta) &=\frac{1}{2}
 \delta\Bigl(\frac{L}{2}\cos\vartheta-|z|\Bigr). \label{EQw0NPlanar}
\end{align}

Carrying out the integrations in Eq.\ (\ref{EQwDiskEffective}) yields
\begin{align}
  w_2^D(z,\vartheta)&=\frac{4\sqrt{\rds{2}-z^2/\sin^2\vartheta}}
                            {\sin\vartheta}\Heavi{\rd\sin\vartheta-|z|},\\
  w_1^D(z,\vartheta)&=\frac{\rd\Heavi{\rd\sin\vartheta-|z|}}
                            {4\sqrt{\rds{2}\sin\vartheta^2-z^2}},
  \label{EQw1dPlanar}\\
  w_0^D(z,\vartheta)
    &=\frac{\Heavi{\rd\sin\vartheta-|z|}}
        {\pi\sqrt{\rds{2}\sin\vartheta^2-z^2}}.\label{EQw0dPlanar}
\end{align}
Comparing Eqs.\ (\ref{EQw1dPlanar}) and (\ref{EQw0dPlanar}) to the
corresponding expressions for needles, Eqs.\ (\ref{EQw1NPlanar}) and
(\ref{EQw0NPlanar}), reveals that the different particle geometries
imprint marked differences in the functional dependence on $z$ and
$\vartheta$.  Clearly this can be traced back to the differences in
the full weight functions for needles [Eqs.\ (\ref{EQw0N}) and
(\ref{EQw1N})] and platelets [Eqs.\ (\ref{EQw1DRosenfeld}) and
(\ref{EQw0DRosenfeld})]. Note that for the present case of vanishing
particle thicknesses, $w_2^D(z,\vartheta)$ has no needle counterpart
to compare with, as the needle surface vanishes.

\begin{align}
  \wv{1}{D}(z,\vartheta)
      &=\frac{z}{8\rd\sin^2\vartheta}
          \sqrt{\frac{\rd^2\sin^2\vartheta-z^2}{\rd^2-z^2}}\ev_z,
\end{align}

\begin{align}
  w_1^{DD}(z,\vartheta,\vartheta') &=
    \begin{cases}
      0 & \text{if }p^2<0\\
      \frac{1}{2}\cos\vartheta' & \text{if }
        p^2\sin(\vartheta-\vartheta')\sin(\vartheta+\vartheta')
          >z^2\sin^2\vartheta'\\
      \frac{1}{\pi
      p}\Bigl(\sqrt{(z^2+p^2)\frac{\sin^2\vartheta'}{\sin^2\vartheta}-p^2}\\
          \quad+p|\cos\vartheta'|\\
          \qquad\times\arcsin\frac{p|\cot\vartheta'|}
              {\sqrt{z^2+\rds{2}\cos^2\vartheta}}\Bigl)
          & \text{else,}
    \end{cases}
\end{align}
where $p=\sqrt{\rds{2}\sin^2\vartheta-z^2}$.

\begin{equation}
  w_2^{DDD}(z,\vartheta,\vartheta',\vartheta'')= 
  \frac{8}{\pi^2}\sqrt{\rds{2}-\frac{z^2}{\sin^2\vartheta}}
   \int_0^{2\pi} \upd\ph'' u(\vartheta,\vartheta',\vartheta'',\ph'')
   \Heavi{\rds{2}\sin^2\vartheta-z^2}
   \label{EQw2DDDplanar}
\end{equation}   
where
\begin{equation}
  u(\vartheta,\vartheta',\vartheta'',\ph'')=
  \begin{cases}
    |\cos\vartheta'\sin\vartheta''\sin\ph''| 
      & \text{if } |k(\vartheta,\vartheta',\vartheta'',\varphi'')|>1\\
   \frac{2}{\pi|\sin\vartheta|}
    \Biggl(\Biggl\{\sin^2\vartheta'
      \Bigl[\sin\vartheta''\cos\vartheta\cos\ph''
        -\sin\vartheta\cos\vartheta''\Bigr]^2 \\
   \qquad +\sin(\vartheta'+\vartheta)\sin(\vartheta'-\vartheta)
        \sin^2\vartheta''\sin^2\ph''
     \Biggr\}^{1/2} \\
   \quad +\sin\vartheta\sin\vartheta''\cos\vartheta'\sin\ph'' 
     \arcsin k(\vartheta,\vartheta',\vartheta'',\ph'')\Biggr)
      & \text{else,}
    \end{cases}
\end{equation}
and
\begin{equation}
  k(\vartheta,\vartheta',\vartheta'',\ph'')=
      \frac{\sin\vartheta\sin\vartheta''\cos\vartheta'\sin\ph''}
        {\sqrt{\sin^2\vartheta'\Bigl[(\sin\vartheta''\cos\ph''\cos\vartheta
          -\sin\vartheta\cos\vartheta'')^2
     +\cos^2\vartheta\sin^2\vartheta''\sin^2\ph''\Bigr]}} 
\end{equation}

For the remaining mixed weight functions we obtain
\begin{align}
  w_2^{DN}(z,\vartheta_D; \vartheta_\needle) 
      &=\frac{4\cot\vartheta_D\cos\vartheta_\needle\sqrt{\rds{2}-z^2}}
          {\rds{2}-z^2/\sin^2\vartheta_D}\Heavi{\rd\sin\vartheta_D-|z|}.
\end{align}

\section{Results}
\label{SECresults}
Already the binary subsystems with platelets (leaving aside the full
ternary mixture), i.e. the sphere-platelet and rod-platelet mixtures,
are expected to display rich bulk and interfacial properties due to
the competition between depletion and orientational order. In order to
find a simple yet demanding test case for the present theory, we
restrict ourselves further to the pure system of platelets. In bulk
this undergoes an isotropic-nematic phase transition, where
$\rho_D(\rv,\Ov)=\rho_D=\rm const$ in the isotropic phase, and
$\rho_D(\rv,\Ov)=\rho_D(\theta)$ in the nematic phase being peaked
around the nematic director that indicates the preferred direction of
alignment of the particles; $\theta$ is the angle between $\Omega$ and
the nematic director. The strength of the nematic order is
conveniently measured via the nematic order parameter,
\begin{align}
  S = \rho_D^{-1} \int \frac{d^2 \Omega}{4\pi}\rho_D(\Ov)
      P_2(\cos\theta),
\end{align}
where $P_2(x)=(3x^2-1)/2$ is the second Legendre polynomial. 

We use the generic form of the free energy, Eq.\
(\ref{EQfexcGeneric}), as a multiple integral over a free energy
density appropriate for pure platelets, $\Phi_D$ as given in Eq.\
(\ref{EQphiD}). This function depends on the weighted densities
$n_2^D$, $n_1^{DD}$, and $n_2^{DDD}$, as defined in Eqs.\
(\ref{EQnnuD}) (with $\nu=2$), (\ref{EQn1DD}), and (\ref{EQn2DDD}),
respectively. The weighted densities in turn are built as convolutions
of the bare platelet one-body density, $\rho(\rv,\Ov)$, with the
weight functions $w_2^D$, $w_1^{DD}$, and $w_2^{DDD}$, as given in
(\ref{EQw2DRosenfeld}), (\ref{EQw1DD}), and (\ref{EQw2DDD}) [and
summarized in Eqs.\ (\ref{EQSUMw2DRosenfeld}), (\ref{EQSUMw1DD}), and
(\ref{EQSUMw2DDD})], respectively. To carry out the actual
calculations, we have started from the corresponding expressions in
planar geometry (see Sec.\ \ref{SECplanarGeometry}), that we have
further (numerically) reduced to $z$-independent quantities, as
appropriate for the spatially homogeneous isotropic and nematic
phases.

Alternatively, our FMT excess free energy functional for hard
platelets can be viewed as being composed of the exact second-order
virial contribution and a further (approximate) third order term that
i) vanishes for configurations without common triple intersection of
three platelets, and ii) is constant for configurations with common
triple intersection; the constant is adjusted to give a reasonable
value for the third virial coefficient. Note that the theory for the
full ternary mixture has a similar structure, but features higher
order (in density) contributions that arise from the finite volume,
and hence finite packing fraction, of the hard spheres.

We minimize the grand potential functional
\begin{equation}
\tilde\Omega(\mu_D,[\rho_D]) = 
F_{\rm id}[\rho_D] + F_{\rm exc}[\rho_D] 
-\mu_D \int d\rv \frac{d\Ov}{4\pi} \rho_D(\rv,\Ov),
\label{EQgrandPotential}
\end{equation}
where the ideal gas contribution is given by
\begin{equation}
  F_{\rm id}[\rho_D] = \int d\rv \frac{d\Ov}{4\pi}
   \rho_D(\rv,\Ov)[\ln(\rho_D(\rv,\Ov)\Lambda_D^3)-1],
\end{equation}
where $\mu_D$ is the chemical potential, the dependence on temperature
and volume has been suppressed in the notation, and we formally set
the thermal wavelength equal to the platelet radius, $\Lambda_D=R_D$,
to fix an arbitrary additive constant to the chemical potential. For
a given value of the (scaled) chemical potential $\mu^\ast=\beta\mu_D$,
we minimized $\tilde\Omega$ with respect to $\rho_D(\Ov)$ in either
phase. Inserting the resulting distribution into Eq.\
(\ref{EQgrandPotential}) yields the grand potential of the system, and
the condition for equality of the grand potential in both phases
locates the phase transition.

The numerical implementation uses free minimization, i.e.\ no
parametrized form of $\rho_D(\theta)$ is assumed a priori. We use an
equidistant grid in $\theta$-space with 100 grid points in the
interval $[0,\pi/2]$. The integration over $\varphi''$ in Eq.\
(\ref{EQw2DDDplanar}) to obtain $w_2^{DDD}$ was done with 200 grid
points in $[0,2\pi]$; this only needs to be performed once at the
start of the calculation. We have used alternatively fixed-point
iteration or simulated annealing to minimize the density profile;
results from both approaches were consistent. Up to 1000 steps were
sufficient to obtain convergence.

\begin{table}
\begin{tabular}{l|l|l|l}
                           & FMT    & simulation & Onsager \\ \hline
$\rho_D^{\rm iso} R_D^3$   & 0.418  & 0.473      & 0.667   \\ \hline
$\rho_D^{\rm nem} R_D^3$   & 0.460  & 0.509      & 0.845   \\ \hline
$\Delta\rho_D R_D^3$       & 0.041  & 0.036      & 0.178   \\ \hline
$S_{\rm nem}$              & 0.492  & 0.37       & 0.781   \\ \hline
\end{tabular}
\caption{Results for the bulk isotropic-nematic phase transition from
  different approaches. Given are values for the (scaled) coexisting
  densities in the isotropic and nematic phase, $\rho_D^{\rm iso}
  R_D^3$ and $\rho_D^{\rm nem} R_D^3$, respectively, the (scaled)
  density jump, $\Delta\rho_D R_D^3=(\rho_D^{\rm nem}-\rho_D^{\rm
  iso})R_D^3$, and the value of the nematic order parameter in the
  coexisting nematic phase, $S_{\rm nem}$, as obtained from FMT, and
  the simulation results by Frenkel and Eppenga~\cite{frenkel82}, as
  well as theoretical results obtained from the Onsager second virial
  functional for hard platelets.}
\label{TABtransitionIN}
\end{table}

\begin{figure}
  \includegraphics[width=\imagewidth]{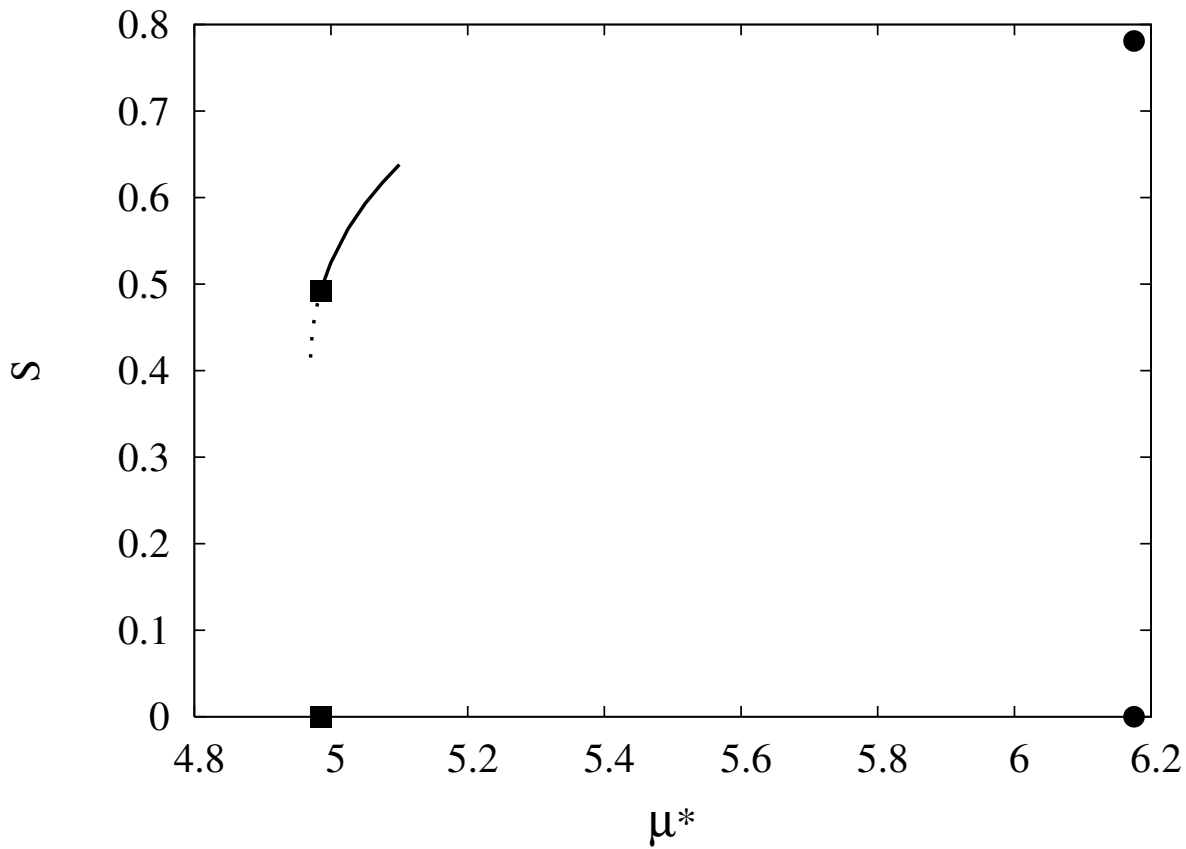}
  \caption{Nematic order parameter $S$ for a bulk system of pure hard
    platelets as a function of the (scaled) chemical potential,
    $\mu^\ast$. Shown are results from FMT in the stable (solid line)
    and metastable (dashed line) nematic. Values in the coexisting
    nematic and isotropic phases, $S_{\rm nem}$ and $S_{\rm iso}=0$,
    are shown as obtained from FMT (squares) and the Onsager-type
    second virial treatment (dots).}
  \label{fig:sofmu}
\end{figure}

\begin{figure}
  \includegraphics[width=\imagewidth]{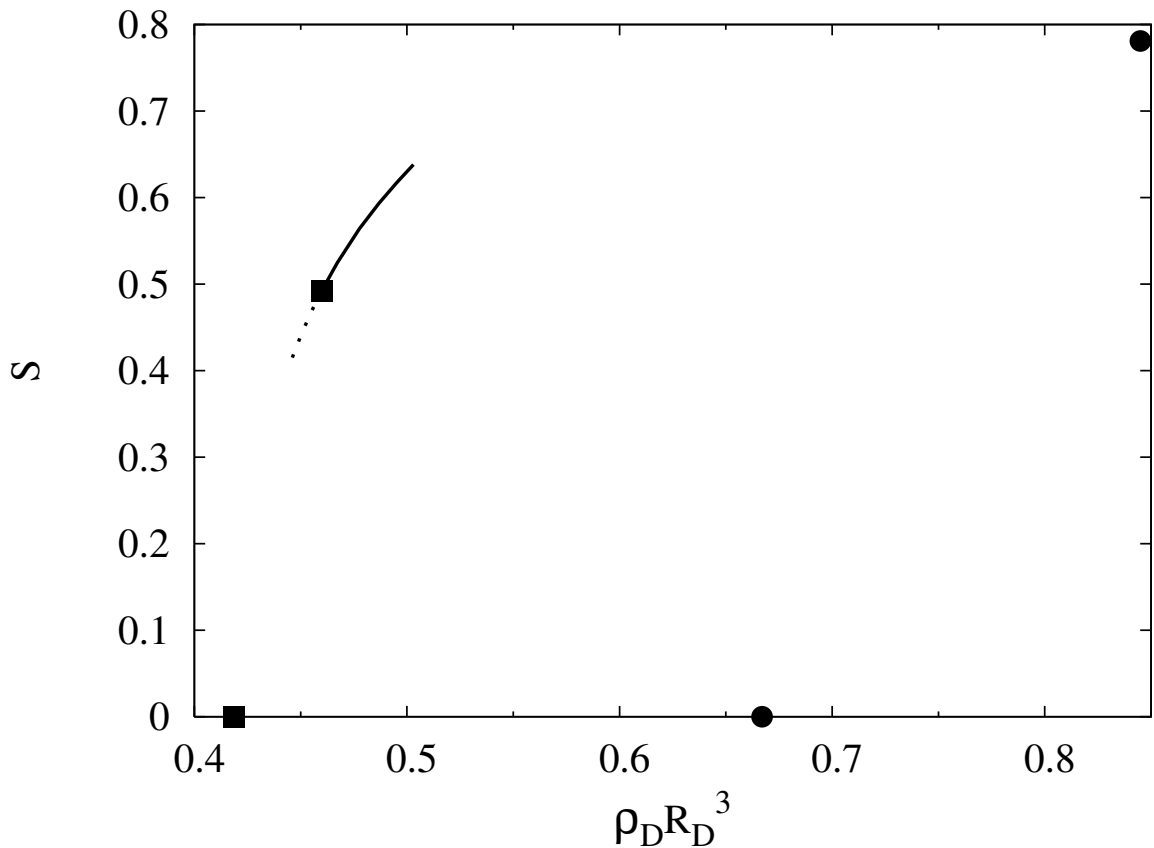}
  \caption{Same as Fig.\ \ref{fig:sofmu}, but as a function of the
    (scaled) density of platelets $\rho_D R_D^3$. Note the
    considerable reduction of the coexistence density gap obtained
    from FMT as compared to the Onsager-like treatment.}
  \label{fig:sofrho}
\end{figure}

The resulting values of the densities in the coexisting isotropic and
nematic phases, $\rho_D^{\rm iso}$ and $\rho_D^{\rm nem}$,
respectively, as well as the value of the order parameter in the
coexisting nematic phase, $S_{\rm nem}$, are in good agreement with
Frenkel and Eppenga's classic simulation results \cite{frenkel82}, see
Tab.\ \ref{TABtransitionIN}.  The FMT values somewhat underestimate
the coexistence densities and give a too high value for $S_{\rm nem}$.
The Onsager second virial functional, on the other hand, overestimates
significantly the coexistence densities as well as $S_{\rm nem}$.
Hence the actual phase transition, as characterized by the jump in
coexistence densities and in order parameter, is weaker than predicted
by the Onsager treatment.
To further illustrate our findings, we plot in Fig.\ \ref{fig:sofmu}
the variation of $S$ with $\mu^\ast$ and in Fig.\ \ref{fig:sofrho} the
variation of $S$ with $\rho_DR_D^3$. 
We also plot the behavior of the metastable nematic branch, i.e.\ for
statepoints where the nematic phase remains locally stable against
small fluctuations, although globally the grand potential for the
isotropic phase is the lower one.
It is apparent that the value of
$S_{\rm nem}$ very sensitively depends on the precise location of the
phase transition. Taking this into account we find the overall the
agreement of the FMT results with simulation data to be very
reasonable, and to give confidence both into applications to
interfacial phenomena of platelets, as well as into the accuracy of
the FMT for the full ternary mixture.

\section{Conclusions}
\label{SECplateconclusions}

In conclusion we have derived a geometry-based density functional
theory for hard body mixtures of spheres, platelets and needles. Both
the needles and the (circular) platelets possess vanishing thickness
and hence constitute the simplest examples of prolate and oblate model
particles, respectively.  Our treatment of the mixture is based on the
so-called deconvolution of the Mayer bond into single-particle
functions which vanish beyond the extent of the particle. The Mayer
bonds are recovered upon convolution of the single-particle functions.
The construction of the full functional relies further on controlling
the third virial level and on Rosenfeld's scaled-particle
treatment. In order to facilitate future applications, like wetting of
planar walls or capillary phenomena in planar slits, we have given
explicit reduced expressions for the relevant quantities in planar
geometry.

In a recent contribution Harnau and Dietrich propose and apply a DFT
for binary platelet-sphere mixtures \cite{harnau04spherePlate}. They
obtain a platelet-sphere functional by starting from the rod-sphere
functional of Refs.\ \cite{schmidt01rsf,brader02rsa}, and in
particular from the explicit expressions for the needle weight
functions in planar and uniaxial geometry
Modifying the definition of the relevant angle between the particle
orientation and the $z$-axis the rod weight functions are taken to
play the role of platelet weight functions. The resulting excess free
energy functional is linear in the platelet density, limiting the
theory to small densities of platelets. Recall that for binary
mixtures where one component (the depletant) is ideal, the absence of
higher than linear order terms in the density distribution of this
component is a good approximation. Examples are the above rod-sphere
mixture and the Asakura-Oosawa model of colloid-polymer mixtures,
where the polymers are described as non-interacting spheres. While in
these cases the pure depletant system is an ideal gas, pure platelets
constitute an interacting system. Hence higher than linear order terms
in the platelet density should only be irrelevant at low platelet
densities. The current theory is in accordance with scaled-particle
theory \cite{barker76} and yields second and third-order contributions
to the excess free energy, but higher order terms are absent. This
behavior can be traced back to the vanishing volume of the platelets.

In the present work we have shown explicitly how our DFT for the
sphere-platelet mixture reduces to the correct low-density limit. In
contrast to the procedure in Ref.\ \cite{harnau04spherePlate}, we have
treated the full three-dimensional problem and have obtained, up to a
defect already present for two-dimensional hard disks, the
deconvolution of the sphere-platelet Mayer bond and hence the
appropriate platelet weight functions. Projecting those to planar and
uniaxial symmetry (appropriate for fluid states at a planar smooth
wall like investigated in Ref.\ \cite{harnau04spherePlate}) reveals
that the expresssions differ markedly from those for sphere-rod
mixtures. This might come as no surprise given the fact that the
genuine shapes of the particles are one of the building blocks of the
geometry-based DFT. An immediate consequence is that the platelet
weight functions, and hence the form of the excess free energy
functional, also differ from the expressions used in Ref.\
\cite{harnau04spherePlate}.

Our successful description of the bulk isotropic-nematic transition of
hard platelets can be viewed as the first genuine FMT treatment of
liquid crystalline ordering in a continuum model (without any
interpolation as e.g.\ in Ref.\ \cite{cinacchi02}).
Possible future applications of our theory include capillary and
wetting phenomena, influence of gravity or other external fields, and
the study of free interfaces between demixed (and possibly liquid
crystalline) phases. Further testing the accuracy of the theory, i.e.\
for the predictions of the bulk direct correlation functions, see
e.g.\ \cite{phuong02}, against results from computer simulations is
clearly desirable.

\acknowledgments Professor Siegfried Dietrich and Dr.\ Ludger Harnau
are acknowledged for useful discussions. We thank Dr.\ Marjolein
Dijkstra and Dr.\ Ren\'e van Roij for stimulating exchange and for
communicating their unpublished results to us. Dr.\ Andrew Archer is
acknowledged for valuable comments on the manuscript. This work is
supported by the SFB-TR6 ``Colloidal dispersions in external fields''
of the German Science Foundation (Deutsche Forschungsgemeinschaft).

\appendix


\end{document}